\documentclass[twocolumn,showpacs,preprintnumbers,amsmath,amssymb,superscriptaddress]{revtex4}

\usepackage{graphicx}
\usepackage{dcolumn}
\usepackage{bm}
\usepackage{tabularx}
\usepackage{ulem}
\usepackage{dsfont}
\usepackage{xfrac}
\newcolumntype{M}{>{\centering\arraybackslash}m{1.85cm}}
\usepackage[export]{adjustbox}
\usepackage{float}
\usepackage{xcolor}   

\graphicspath{{figures/}}
\usepackage{hyperref}
\hypersetup{
    colorlinks=true, 
    linktoc=all,     
    linkcolor=blue,  
}

\makeatletter
\newcommand{\colorcaption}[2][]{%
  \begingroup%
  \renewcommand{\@caption@fignum@sep}{ (Color online). }%
  \caption[#1]{#2}%
  \endgroup%
}
\makeatother
\begin{document}
\title{ \bf\textit{Ab initio} no-core shell model description of $^{10-14}$C isotopes}

\author{ Priyanka Choudhary}
\address{Department of Physics, Indian Institute of Technology Roorkee, Roorkee 247667, India}

\author{Praveen C. Srivastava}
\address{Department of Physics, Indian Institute of Technology Roorkee, Roorkee 247667, India}

\author{Michael Gennari}
\affiliation{University of Victoria, 3800 Finnerty Road, Vicotria, British Columbia V8P 5C2, Canada}
\affiliation{TRIUMF, 4004 Wesbrook Mall, Vancouver, British Columbia V6T 2A3, Canada}

\author{Petr Navr{\'a}til}
\affiliation{TRIUMF, 4004 Wesbrook Mall, Vancouver, British Columbia V6T 2A3, Canada}

\date{\hfill \today}


\begin{abstract}
We present a systematic study of the $^{10-14}\text{C}$ isotopes within the \textit{ab initio} no--core shell model theory. We apply four different realistic nucleon--nucleon (NN) interactions: (i) the charge--dependent Bonn 2000 (CDB2K) potential (ii) the inside non--local outside Yukawa (INOY) potential (iii) the next--to--next--to--next--to--leading order (N\textsuperscript{3}LO) potential, and (iv) the optimized next--to--next--to--leading order (N\textsuperscript{2}LO\textsubscript{$opt$}) potential. We report the low--lying energy spectra of  both positive and negative parity states for the $^{10-14}\text{C}$ isotopes  and investigate the level structures. We also calculate electromagnetic properties such as transition strengths, quadrupole and magnetic moments.  The dependence of point--proton radii on the harmonic oscillator frequency and basis space is shown. We present calculations of the translation invariant one--body density matrix in the no--core shell model and discuss isotopic trends in the density distribution. The maximum basis space reached is $10 \hbar \Omega$ for $^{10}\text{C}$ and $8 \hbar \Omega$ for $^{11-14}\text{C}$, with a maximum M--scheme dimension of $1.3 \times 10^{9}$ for $^{10}\text{C}$.   We found that while the INOY interaction gives the best description of the ground state energies, the N$^3$LO interaction best reproduces the point--proton radii.
\end{abstract}

\maketitle
\newpage

\section{Introduction}

Over the last few decades, the main goal of nuclear theorists has been to explain the structure of the atomic nucleus from \textit{ab initio} theory \cite{NCSM2013}. Several \textit{ab initio} approaches, in which nucleons interact via realistic inter--nucleon potentials, have been developed to study the properties of nuclei. Examples of approaches utilized to solve the many--body problem include the in--medium similarity renormalization group (IM--SRG)~\cite{IMSRG4,IMSRG1,IMSRG2,IMSRG3,IMSRG} approach, the coupled--cluster effective interaction (CCEI)~\cite{CC,CC1,CC2,CC3} approach, the quantum Monte Carlo (QMC) method~\cite{QMC0,QMC1,QMC2}, the no--core shell model (NCSM) \cite{Phys.Rev.C502841(1994),Phys.Rev.C542986(1996),Phys.Rev.C573119(1998),PRC1906(1999),PRC62,2009,MVS2009,NCSM2013,stetcu1,stetcu2} approach, many--body perturbation theory (MBPT)~\cite{MBPT}, self--consistent Green's function (SCGF)~\cite{SCGF1,SCGF2} method, \textit {etc.} In addition to the complexity of quantum many--body methods, it is challenging to develop the inter--nucleon interactions from first principles. Realistic potentials are frequently built on the basis of either meson--exchange theory or the underlying symmetries of quantum chromodynamics (QCD)~\cite{QCD}, e.g. via chiral effective field theory ($\chi$EFT)~\cite{RMP,EFT1,EFT2,EFT3}. In the past few years, several works have been dedicated to investigate the carbon isotopes using \textit{ab initio} many--body approaches, with various two--nucleon (NN) and three--nucleon (3N) interactions employed. While we are specifically interested in application of the NCSM theory, we review several of these works here. 

The binding energy and reduced electric quadrupole transition strength $B(E2)$ of $^{10}\text{C}$ were calculated with the Argonne V8' (AV8')~\cite{av8'} and CDB2K~\cite{A=10} NN potentials. At the time, NCSM calculations of $^{10}\text{C}$ with the CDB2K interaction were only possible in a model space of $8 \hbar \Omega$. In more recent studies, it has become possible to extend the size of the basis space to $10 \hbar \Omega$~\cite{C.Forssen}. Additional work has been dedicated to the description of $^{12}\text{C}$, as it has the highest isotopic abundance in nature and is formed in stellar nucleosynthesis reactions. Structurally, this nucleus poses a significant challenge for shell model approaches due to its triple $\alpha$--state, i.e. the Hoyle state. In Refs.~\cite{P.Navratil,A.C.Hayes,PRL99}, $^{12}\text{C}$ was studied using NCSM theory with the AV8' and AV8'+Tucson--Melbourne (TM'99)~\cite{TM99} potentials at basis spaces of $4 \hbar \Omega$, in addition to the chiral N\textsuperscript{3}LO and CDB2K potentials at $6 \hbar \Omega$ basis spaces. In Ref.~\cite{2021}, the authors have applied a semilocal momentum space regularized chiral interaction up to third order (N\textsuperscript{2}LO) to study the properties of $^{12}\text{C}$. The binding energy increases slightly with N\textsuperscript{2}LO as compared to NLO. With the addition of 3N forces, overbinding of the states are observed.

The N\textsuperscript{3}LO potential was employed within the NCSM to study of energy spectra for $^{13}\text{C}$ in Ref.~\cite{PRL99}, utilizing a $6\hbar \Omega$ basis space. In this work, an argument for the inclusion of 3N interactions is made, though even with 3N forces some states are not adequately matched to experiment. These works establish the success of \textit{ab initio} methods in describing the carbon isotopes, however they also indicate that larger basis spaces are required in order to obtain converged results. In addition to the aforementioned studies, Forss{\'e}n \textit{et al.}~\cite{C.Forssen} investigated the low--lying states of even carbon isotopes in the range of $A = 10-20$ using the CDB2K potential within the NCSM approach. The dependence of the ground state (g.s) and the excited $2^{+}_{1}$ energies, the $B(E2; 2^{+}_{1} \rightarrow 0^{+}_{1})$ transition rates and the $2^{+}_{1}$ quadrupole moments on the NCSM basis expansion parameters was discussed in this work.

In Ref.~\cite{A.M.Shirokov}, the binding energy of $^{12}\text{C}$ was computed in the NCSM using the Daejeon16 NN interaction with an $N_\mathrm{max} = 8$ basis space, which was compared to an $N_\mathrm{max} = 10$ calculation using the J--matrix inverse scattering potential (JISP16). In addition, the $\hbar\Omega$ dependence of $^{12}\text{C}$ observables, such as the point--proton root--mean--square (r.m.s.) radii ($r_p$), is contrasted between the two different interactions. In general, it was found that $r_p$ is relatively sensitive to the NCSM expansion parameters $\hbar\Omega$ and $N_\mathrm{max}$. The study found that \textit{ab initio} calculations utilizing the Daejeon16 interaction better described the binding energy and $r_p$ of $^{12}\text{C}$ than results obtained with the JISP16 interaction. So, calculation with the Daejeon16 NN interaction indicates that a reasonable description of $^{12}\text{C}$ can be obtained with only two-nucleon interaction. In Ref.~\cite{K.Kaki}, the r.m.s. radius of protons, neutrons, matter, and the charge distribution for $^{8-22}\text{C}$ were obtained using relativistic mean--field calculations. For the smaller carbon isotopes, a larger proton r.m.s. radius was observed due to the strong effects of Coulomb repulsion between the protons. 

Experiment performed at GSI,  Darmstadt~\cite{GSI} to investigate proton radii for the carbon isotopic chain $^{12-19}\text{C}$, using the technique of charge--changing cross--section measurements for the first time. In this study, the authors observed a rapidly increasing neutron skin in the larger isotopes $^{15-19}\text{C}$, which suggests the formation of neutron halos. Further, \textit{ab initio} CC calculations are performed using two different potentials, the N\textsuperscript{2}LO\textsubscript{sat} \cite{N2LOSat} and N\textsuperscript{2}LO\textsubscript{$opt$} \cite{N2LOopt,optimizedn2lo} interactions, which are then compared to the experimental results. The CC results obtained with the N\textsuperscript{2}LO\textsubscript{sat} interaction provide good agreement with the data in comparison to results obtained with the N\textsuperscript{2}LO\textsubscript{$opt$} interaction.  Yet another experiment, detailed in Ref.~\cite{RCNP}, at the Research Center for Nuclear Physics (RCNP) at Osaka University utilized the same charge--changing cross--section technique for measuring the proton r.m.s. radii for $^{12-16}\text{C}$, further expanding the abundance of precision carbon measurements for theoretical comparison.

In recent years, experimental evidence for a proton sub--shell closure at $Z = 6$ was discovered, in particular in neutron--rich carbon isotopes~\cite{Z=6}. The authors presented a systematic study of $^{13-20}\text{C}$ point--proton radii, electric quadrupole transition rates $B(E2; 2_1^+ \rightarrow 0_{g.s}^+)$, and atomic mass data. From this information, the authors are able to deduce a sub--shell magic number at $Z = 6$. The experimental observations were further supported by \textit{ab initio} CC theory calculations with the SRG--evolved NN+3N chiral and the N\textsuperscript{2}LO\textsubscript{sat} interactions. In Ref.~\cite{M.Stanoiu}, the isotopes $^{17-20}\text{C}$ are studied and the author found evidence that there is no shell closure at $N = 14$, which is anticipated in carbon due to its appearance in the oxygen isotopes. In Ref.~\cite{Jansen}, an effective valence--space shell model interaction is derived from \textit{ab initio} CC theory and implemented in calculations of the carbon isotopes $^{17-22}\text{C}$. Good agreement between CCEI results and experiment is found, with further indication of a weaker shell closure at $N = 14$. In Ref. \cite{YSOX}, the energy level structures for $^{12-20}$C have been studied using the shell model with a newly constructed monopole based universal interaction.

In this paper, we have employed the \textit{ab initio} NCSM theory in description of the nuclear structure properties of $^{10-14}\text{C}$ isotopes using the INOY, CDB2K, N\textsuperscript{3}LO and N\textsuperscript{2}LO\textsubscript{$opt$} realistic NN interactions. We have determined the g.s. energies, the positive and negative parity excitation spectra, the reduced transition probabilities, the quadrupole moments, the magnetic moments, and the point--proton radii of the carbon isotopes. In terms of the basis space, the maximum basis size reached is 10$\hbar \Omega$ for $^{10}\text{C}$ and 8$\hbar \Omega$ for $^{11-14}\text{C}$.  In addition, we provide a comparison between \textit{ab initio} results and experiment.  For the first time, a comparison of the local and nonlocal g.s. nuclear densities of the carbon isotopes are presented. Using the CDB2K interaction, we study the convergence trends of the isotopes with increasing $N_\mathrm{max}$.

In Sec.~\ref{sec:ncsm_formalism}, we  explain the \textit{ab initio} NCSM formalism. In Sec.~\ref{sec:effective_interactions}, we  introduce the realistic interactions employed in construction of the NCSM Hamiltonian. Then, in Sec.~\ref{sec:spectra_results}, we present and discuss the results of the energy spectra for $^{10-14}\text{C}$, further comparing them with  experimental data. In Sec.~\ref{sec:em_results}, we discuss the electromagnetic properties focussing on the $B(M1)$ and $B(E2)$ transition strengths. In Sec.~\ref{sec:protonradii_results}, we report the point--proton radii for the g.s. of $^{12}\text{C}$ utilizing the INOY and N\textsuperscript{3}LO interactions. Lastly, in Sec.~\ref{sec:ncsm_densities}, we show results for the one--body density matrix calculated in the NCSM. We then draw our conclusions in Sec.~\ref{sec:conclusions}.

\section{No-CORE SHELL MODEL FORMALISM}\label{sec:ncsm_formalism}

The NCSM~\cite{2009,NCSM2013} is a non--relativistic many--body theory suited to describing low--lying bound and resonant states of $s$--, $p$-- and light $sd$--shell nuclei. In the NCSM, all nucleons are treated as point--like and are considered to be active degrees of freedom. Unlike traditional nuclear shell models, there is no concept of an inert core. The translation invariance of observables, angular momentum, and the parity of a given system are all conserved in this approach.

Consider an $A$--nucleon system interacting through a realistic nuclear interaction. In general, one considers realistic NN and 3N potentials typically derived within the $\chi$EFT. In the present work, we have considered only realistic NN interactions and took advantage of an access to a larger basis space. The nuclear Hamiltonian is then given by:

\begin{equation}\label{eq:H_A}
H_A = T_{\rm rel} + V = \frac{1}{A}\sum_{i<j}^{A}\frac{(\vec{p}_i-\vec{p}_j)^2}{2m} + \sum_{i<j}^{A}V^{\text{NN}}_{ij} \ ,
\end{equation}

where, $T_{\rm rel}$ is the relative kinetic energy of the nucleons, $m$ is the nucleon mass and $V^{\rm NN}_{ij}$ corresponds to the realistic two--body interaction containing both strong and electromagnetic parts.

Solving the nuclear Hamiltonian requires a choice of basis. In the NCSM, we make use of a finite (but large) set of antisymmetrized HO many--body basis states. The HO many--body states preserve the symmetry properties of the Hamiltonian, i.e. they conserve angular momentum, parity, and isospin. The most desired property, however, is that the HO basis preserves translation symmetry of the $A$--nucleon system even when utilizing single--nucleon coordinates. The basis truncation parameter $N_\mathrm{max}$ defines the total number of allowed oscillator quanta above the lowest Pauli configuration of the $A$--nucleon system. The HO many--body expansion in the Cartesian coordinate Slater determinant (SD) basis is given by:

\begin{equation}
\big\langle \vec{r}_1 \cdots \vec{r}_A \big\vert \Psi^{J^{\pi}T}_A \big\rangle{}_{\text{SD}} = \sum_{N=0}^{N_\mathrm{max}} \sum_{\alpha} c_{N\alpha}^{(\text{SD})} \big\langle \vec{r}_1 \cdots \vec{r}_A \big\vert \Phi_{N\alpha}^{A} \big\rangle{}_{\text{SD}}, 
\end{equation}

where, $\big\vert \Phi_{N\alpha}^A \big\rangle$ corresponds to the HO many--body states. In the case of hard potentials, i.e. those which produce strong short--range correlations, one often needs large $N_\mathrm{max}$ in order to reach many--body convergence. However, basis space growth is exponential with respect to $N_\mathrm{max}$ and we quickly reach a computational ceiling. Thus, at the moment it is necessary to employ renormalization methods which weaken the short--range repulsion of nuclear interactions, allowing for reduced correlations in the many--body calculation. This `softening' of the interactions has been shown to accelerate many--body convergence. Such renormalization schemes include the Okubo--Lee--Suzuki (OLS) method~\cite{Prog.Theor.Phys.12,Prog.Theor.Phys.,Prog.Theor.Phys.68,Prog.Theor.Phys.92} and the similarity renormalization group (SRG) approach~\cite{Phys.Rev.C.75.061001,Annals.Phys.506.77}. In the present work we apply the OLS approach.

In order to facilitate the expansion in the HO basis, and the derivation of the OLS effective interactions in truncated basis spaces, we add the center--of--mass (c.m.) Hamiltonian for an A--body HO to the original Hamiltonian in Eq.~(\ref{eq:H_A}), which is given by:

\begin{equation}\label{COM}
H_{\text{c.m.}} = T_{\text{c.m.}} +U_{\text{c.m.}} \ ,
\end{equation}

where

\begin{equation}
U_{\text{c.m.}} = \frac{1}{2}Am{\Omega}^2{\vec{R}}^2, \qquad \vec{R} = \frac{1}{A} \sum _{i=1}^{A} \vec{r}_i \ .
\end{equation}

The modified Hamiltonian then has a frequency dependent form

\begin{equation}\label{H^w}
\begin{split}
& H^{\Omega}_{A} = H_{A} + H_{\text{c.m.}} = \sum _{i=1}^{A} h_{i} + \sum _{i<j}^{A}V^{\Omega,A}_{ij} \\
& \ = \sum_{i=1}^{A} \bigg[ \frac{\vec{p}_i^2}{2m} + \frac{1}{2}m{\Omega}^2 \vec{r}_i^2 \bigg] + \sum_{i<j}^{A} \bigg[ V_{ij}^{\text{NN}} - \frac{m {\Omega}^2}{2A} {(\vec{r}_i - \vec{r}_j)}^2 \bigg] \ ,
\end{split}
\end{equation}

where $\Omega$, the HO frequency, is a variational parameter in the NCSM. Since the initial Hamiltonian of Eq.~(\ref{eq:H_A}) is translation invariant, addition of the c.m. Hamiltonian will not impact the intrinsic properties of the Hamiltonian. With $H_A^{\Omega}$ in this form, it is possible to determine the effective interactions for a given model space. 

We first separate the infinite HO basis into two parts by utilizing the projectors $P$ and $Q$, defined such that $P+Q={1}$. The $P$--space contains all HO basis states up to the truncation parameter $N_\mathrm{max}$, and the $Q$--space contains all excluded states. Thus NCSM calculations are performed solely in the $P$--space.  The effective interaction is then obtained by applying the OLS unitary transformation on the Hamiltonian \ref{H^w}. As a result of the OLS transformation, this new Hamiltonian contains induced terms up to the A--body level. Moreover, for an arbitrary $A$--nucleon system, one generally needs $A$--body effective interactions to exactly reproduce the original Hamiltonian. However, since we know that the two--body potential is the dominant contribution in the interaction, it is a reasonable approximation to keep only the two--body effective interaction terms.
Finally, by subtracting out the c.m. Hamiltonian and adding the Lawson projection term~\cite{Lawson} (which shifts out spurious c.m. excitations), the effective Hamiltonian takes the form:

\begin{equation}
\begin{split}\label{eq:H_eff_OLS}
& H_{A, \rm eff}^{\Omega} = P \Bigg\{ \sum_{i<j}^{A} \bigg[ \frac{{(\vec{p}_i - \vec{p}_j)}^2}{2mA} + \frac{m {\Omega}^2}{2A} {(\vec{r}_i - \vec{r}_j)}^2 \bigg] \\
&  + \sum_{i<j}^{A} \bigg[ V^{\rm NN}_{ij} - \frac{m {\Omega}^2}{2A}{(\vec{r}_i - \vec{r}_j)}^2 \bigg]_{\text{eff.}} 
 \hspace{-0.3cm} + \beta \bigg( H_{c.m.} - \frac{3}{2}\hbar\Omega \bigg) \Bigg\} P \ .
\end{split}
\end{equation}

 The resulting Hamiltonain of Eq. (6) is hence dependent
on the HO frequency $\hbar$$\Omega$, the basis truncation parameter $N_\mathrm{max}$ and the parameter $\beta$. In our calculations, we fix $\beta$=10. Since $\beta$ is fixed, the Hamiltonian depends on only two parameters: $\hbar$$\Omega$ and $N_\mathrm{max}$.
For $N_\mathrm{max} \rightarrow \infty$, the effective Hamiltonian approaches the non--renormalized Hamiltonian and the results are exactly reproduced.

\section{Effective inter--nucleon interactions}\label{sec:effective_interactions}

In this work, we have chosen four different interactions to analyse many--body results: (i) the inside nonlocal outside Yukawa (INOY) potential ~\cite{INOY,nonlocal,Doleschall} (ii) the charge--dependent Bonn 2000 (CDB2K) potential~\cite{Machleidt1,Machleidt2,Machleidt3,Machleidt4}  (iii) the chiral next--to--next--to--next--to--leading order (N\textsuperscript{3}LO) potential~\cite{QCD,Machleidt} and (iv) the optimized chiral next--to--next--to--leading order (N\textsuperscript{2}LO\textsubscript{$opt$}) potential~\cite{N2LOopt,optimizedn2lo}. 

The INOY NN potential is formulated such that the short--range physics (within some radius $r \sim 1$ fm) is properly nonlocal, and the interaction is treated as local outside this radius. The short--range nonlocality of the INOY interaction is the outcome of the internal structure of the nucleons, while the long--range local component is a Yukawa--type potential, similar to that of the Argonne V18 potential discussed in Ref.~\cite{av18}. The ranges of both the local and nonlocal features are adjustable as the interaction is derived in coordinate space. The form of the INOY NN interaction is given by:
\begin{equation*}
V_{l l'}^{\rm full}(r,r') = \delta(r - r') F_{l l'}^{\rm cut}(r) V_{l l'}^{\rm Yukawa}(r) + W_{l l'}(r, r') \ .
\end{equation*}
The first term in this expression is the local interaction, characterized by the $V_{l l'}^{\rm Yukawa}(r)$ Yukawa tail. $F_{l l'}^{\rm cut}(r)$ is the cutoff function applied to the interaction, which is defined in coordinate space as:
\begin{equation*}
F^{\rm cut}_{ll'}(r) = \theta(r - R_{ll'}) \bigg( 1- e^{-\big[ \alpha_{ll'}(r - R_{ll'}) \big]^{2}} \bigg) \ , 
\end{equation*}
where,
\begin{equation*}
\theta(r - R_{l l'}) = \begin{cases} 1 & \text{for} \ r > R_{l l'} \\ 0 & \text{for} \ r \leq R_{l l'} \end{cases} \ .
\end{equation*}
The second term $W_{l l'}(r, r')$ is the nonlocal component of the interaction. When inter--nucleon interactions are used to describe atomic nuclei, it becomes necessary to incorporate many--nucleon forces, e.g. three--body forces, for a complete description. 
In the case of the INOY interaction, there is no immediate need to include 3N forces due to the short--range nonlocal character of the potential which captures some of the intrinsic structure of the nucleon. This interaction is capable of reproducing the properties of $^3\text{H}$ and $^3\text{He}$ with a high degree of precision.

The CDB2K NN potential is based on meson--exchange theory. The mesons which exist below the threshold of the nucleon mass, e.g. the $\pi^{\pm,0}$, $\eta$, $\rho$, $\omega$ and the two scalar--isoscalar $\sigma$ bosons, are included. The construction of the potential is based on analysis of the covariant one--boson--exchange Feynman diagrams, the nonlocal character of which is particularly valuable in solving the problem of underbinding in nuclei.

The chiral N\textsuperscript{3}LO interaction is obtained from a formulation of inter--nucleon forces based on EFT for low--energy QCD. By studying the low--energy symmetry of QCD, i.e. chiral symmetry, it is possible to perturbatively construct two-- and many--nucleon interactions. Derived in the formalism of chiral perturbation theory ($\chi$PT), the chiral orders are represented in powers of $Q/\Lambda_{\chi}$, where $Q$ is the low--momentum scale being probed, usually on the order of the pion mass, and $\Lambda_{\chi}$ corresponds to the chiral symmetry breaking scale $\sim 1$ GeV (the mass of the nucleon). The N\textsuperscript{3}LO potential of Entem and Machleidt is computed at fourth order in the $\chi$PT expansion and has a precision comparable to one of the many high--quality phenomenological potentials, yielding a $\chi^2/\text{datum} \approx 1$ for data up to $290$ MeV.

Lastly, the optimized chiral N\textsuperscript{2}LO\textsubscript{$opt$} interaction is a chiral potential at next--to--next--to--leading order in the $\chi$PT expansion which has been optimized via the POUNDerS (Practical Optimization Using No Derivatives for the Sum of Squares) algorithm up to a laboratory energy of $125$ MeV. The three pion--nucleon couplings ($c_1$, $c_3$ and $c_4$) which emerge in $\chi$PT, as well as the $11$ partial wave contact parameters ($C$ and $\tilde{C}$), are varied and optimized using POUNDerS. The axial--vector coupling constant $g_A$, the pion decay constant $F_{\pi}$ and all particle masses remain as constants. Comparatively, the N\textsuperscript{2}LO\textsubscript{$opt$} provides substantially better performance than the standard N\textsuperscript{2}LO interaction, and is capable of reproducing nuclear properties such as the binding energies and radii of $A=3,4$ nuclei, the position of the neutron drip line in the oxygen chain, shell--closures in the calcium isotopes and the neutron matter equation of state at subsaturation densities, all without the addition of 3N forces. The standard N\textsuperscript{2}LO produces a $\chi^2/\text{datum} \approx 10$ below $125$ MeV, whereas the N\textsuperscript{2}LO\textsubscript{$opt$} interaction produces a $\chi^2/\text{datum} \approx 1$.

For the present calculations using NCSM theory, we have applied the OLS renormalization in calculations with the INOY, CDB2K, and N$^3$LO NN interactions (Hamiltonian given in Eq.~(\ref{eq:H_eff_OLS})), while no renormalization was used for the N\textsuperscript{2}LO\textsubscript{$opt$} interaction (Hamiltonian given in Eq.~(\ref{eq:H_A})). We utilized the pAntoine code~\cite{pAntoine1,pAntoine11,pAntoine2} for the \textit{ab~initio} calculations.  Our group has recently reported NCSM results in similar studies of the boron, nitrogen, oxygen and  fluorine isotopes in Refs.~\cite{pc1,arch1,arch2}. 

\section{Results and Discussions}\label{sec:spectra_results}
\vspace{-0.5cm}
\begin{table}[h]
\centering
\caption{\label{tbl:dimension_isotopes} The dimensions corresponding to different model spaces for the $A = 10-14$ carbon isotopes are presented. The Hamiltonian dimensions for which we have performed NCSM calculations are shown in blue.}
\begin{adjustbox}{width=0.48\textwidth}
	\begin{tabular}{cccccccccccc}
		\hline
		\hline\\
		\hspace{-0.1cm}$N_{\mbox{max}}$  &  $^{10}$C  & $^{11}$C  &  $^{12}$C  &$^{13}$C  &  $^{14}$C \\
		\hline \vspace{-2.8mm}\\
		0 & {\color{blue} 51}  & {\color{blue} 62}  &{ \color{blue} 51} &{ \color{blue} 21}& {\color{blue} 5} \\
		2 & { \color{blue} $ 1.0\times10^{4}$ } & { \color{blue} $ 1.6\times10^{4}$ } &{ \color{blue} $ 1.8\times10^{4}$ } & { \color{blue} $ 1.2\times10^{4}$ } &{ \color{blue}$5.8\times10^{3}$} \\
		4 & { \color{blue} $ 4.3\times10^{5}$ }  & { \color{blue} $ 8.1\times10^{5}$ }& { \color{blue} $ 1.1\times10^{6}$ }  & { \color{blue} $ 1.1\times10^{6}$ } &{  \color{blue}$ 7.3\times10^{5}$} \\
		6 & {\color{blue} $9.2\times10^{6}$} & {\color{blue} $2.0\times10^{7}$} & {\color{blue} $3.3\times10^{7}$} & {\color{blue} $3.8\times10^{7}$} &{ \color{blue} $3.4\times10^{7}$} \\
		8 & {\color{blue} $1.3\times10^{8}$} & {\color{blue} $3.2\times10^{8}$}& {\color{blue} $5.9\times10^{8}$} & {\color{blue} $8.2\times10^{8}$} &{ \color{blue}$ 8.7\times10^{8}$} \\
		10 & {\color{blue} $1.3\times10^{9}$} &  $3.7\times10^{9}$&  $7.8\times10^{9}$ & $1.2\times10^{10}$ &{ $1.5\times10^{10}$} \\
		\hline \hline
	\end{tabular}
\end{adjustbox}
\end{table}
\begin{figure*}
	\includegraphics[width=8.75cm]{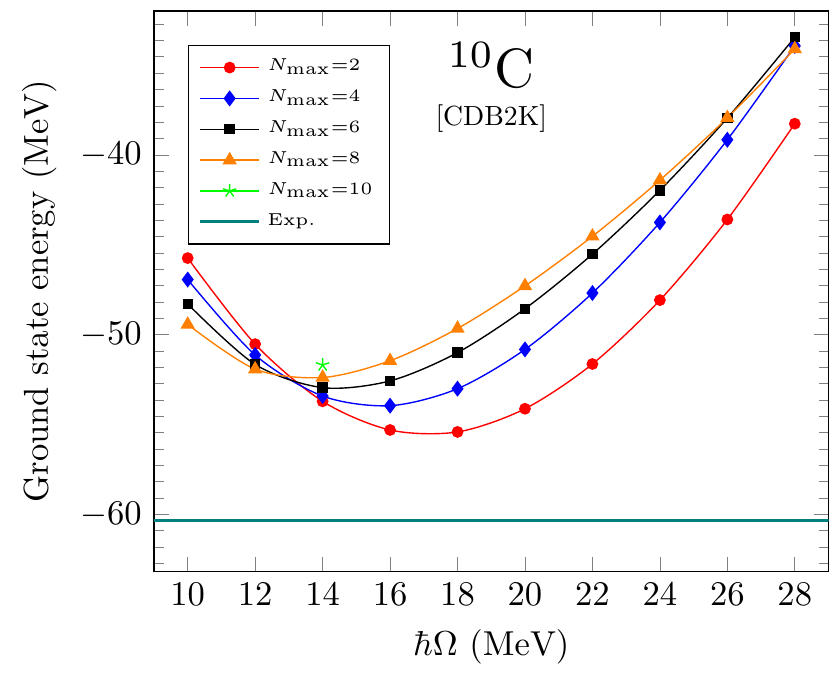}
	\includegraphics[width=8.75cm]{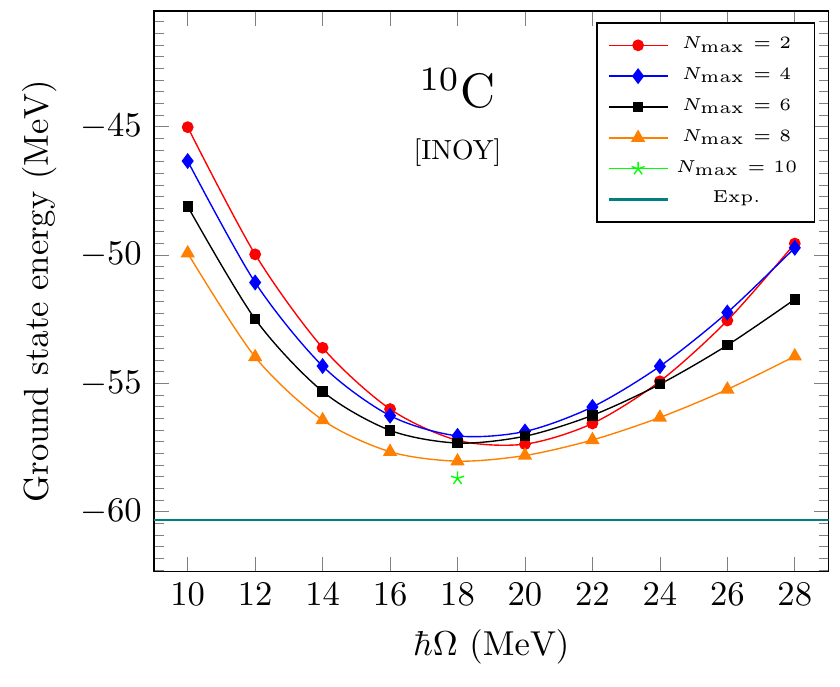}
	\includegraphics[width=8.75cm]{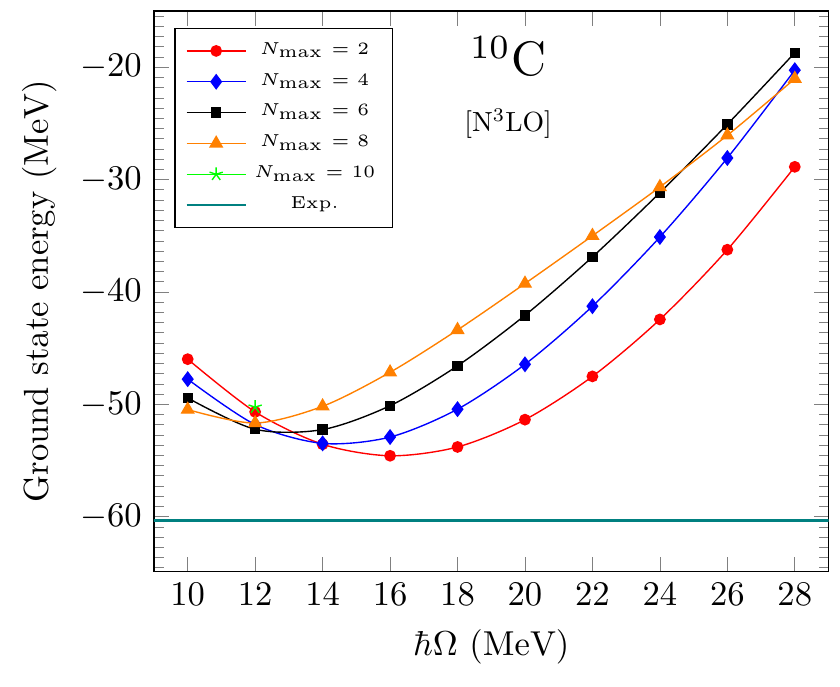}
	\includegraphics[width=8.75cm]{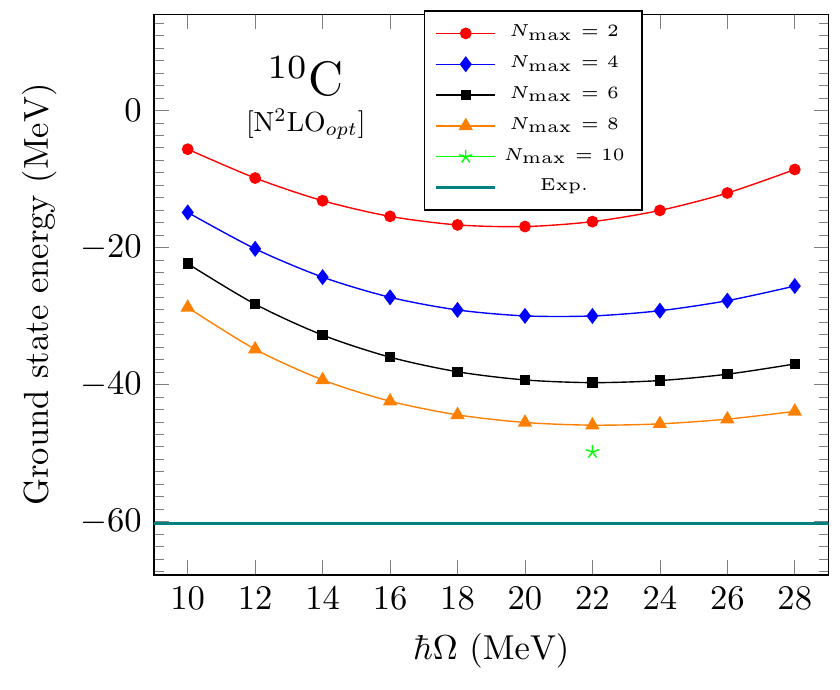}
	\caption{\label{fig:10C_Egs} Dependence of the g.s. energy for $^{10}\text{C}$ on HO frequency ranging from 10--28 MeV is presented. NCSM results are obtained using the CDB2K, INOY,  N\textsuperscript{3}LO and N\textsuperscript{2}LO\textsubscript{$opt$} interactions. We show NCSM calculations for $N_\mathrm{max} = 2-8$ and include the $N_\mathrm{max}=10$ result evaluated at the minima corresponding to $N_\mathrm{max} = 8$, of the respective interactions. The experimental g.s energy for $^{10}\text{C}$ is indicated by a horizontal line~\cite{NNDC}.}
\end{figure*}
In this section we have presented NCSM results for the spectra of $^{10-14}\text{C}$. The Hamiltonian dimensions calculated in M--scheme for $^{10-14}\text{C}$ are shown in Table~\ref{tbl:dimension_isotopes} for model space sizes corresponding to $N_\mathrm{max} = 0-10$. The Hamiltonian dimension poses a significant challenge in nuclear theory due to its exponential behavior with respect to the truncation parameter $N_\mathrm{max}$, and so advances in computational power are necessary for the extension of the NCSM calculations to heavier systems. With the available computational resources, we have successfully reached basis spaces up to $N_\mathrm{max} = 10$ for $^{10}\text{C}$ and $N_\mathrm{max} = 8$ for $^{11-14}\text{C}$, the corresponding dimensions  are highlighted in blue in Table~\ref{tbl:dimension_isotopes}. 
We have reached maximum dimension of $1.3 \times 10^{9}$ for $^{10}\text{C}$.

Analysis of the NCSM calculations is generally a two--step process. In the initial step, the minimum of g.s. energy with respect to the oscillator frequency is determined. The convergence of the g.s. energy of $^{10}\text{C}$ with respect to the HO frequency parameter is shown in Fig.~\ref{fig:10C_Egs}, for $N_\mathrm{max}=2-8$ basis spaces corresponding to the four aforementioned interactions. 
The oscillator frequency is varied on the interval $10-28$ MeV. It is clear that as we increase the basis truncation parameter $N_\mathrm{max}$ the dependence of the g.s. energy on the oscillator frequency is reduced, as expected. It should be noticed that due to the application of the OLS renormalization, the CDB2K, INOY, N\textsuperscript{3}LO calculations are not variational.  We observe, however, that the INOY results behave variational-like for $N_{\rm max}{>}2$ spaces.  Inclusion of the three--body effective interaction does not change the non--variational nature of the method; it contributes either positively or negatively to the binding energy. The N\textsuperscript{2}LO\textsubscript{$opt$} calculations are, on the other hand, variational as no Hamiltonian renormalization is applied. 
 
We have extracted the HO frequency corresponding to minimal g.s. energy in the $N_\mathrm{max} = 8$ space as optimal and used it for the g.s. energy in the $N_\mathrm{max}{=}10$ space. The optimal frequencies for $^{10}\text{C}$ are then $\hbar \Omega{=}14, 18, 12$, and 22 MeV for the CDB2K, INOY, N\textsuperscript{3}LO and N\textsuperscript{2}LO\textsubscript{$opt$} interactions, respectively.
Similarly, we have determined the respective optimal frequencies for other carbon isotopes. In general, we obtain lower optimal frequencies for the CDB2K and N\textsuperscript{3}LO interactions, while we observe larger optimal frequencies for the INOY and N\textsuperscript{2}LO\textsubscript{$opt$} interactions. It is clear that the optimal frequency of the HO expansion is highly sensitive to the nuclear interaction employed. Once the optimal frequency is determined for a particular nucleus and interaction, the next step is to determine  the energy spectra.

\begin{figure*}
	\includegraphics[width=17cm]{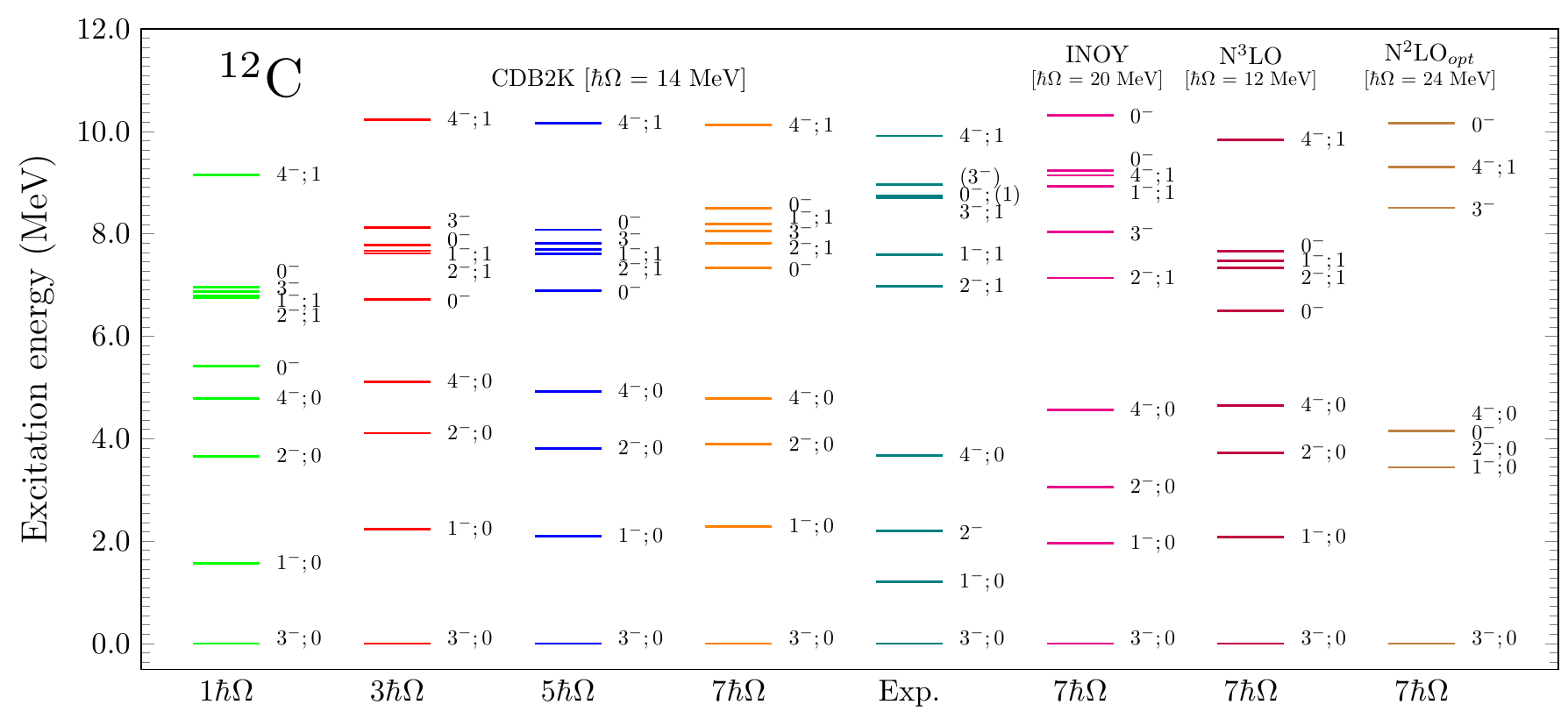}
	\includegraphics[width=17cm]{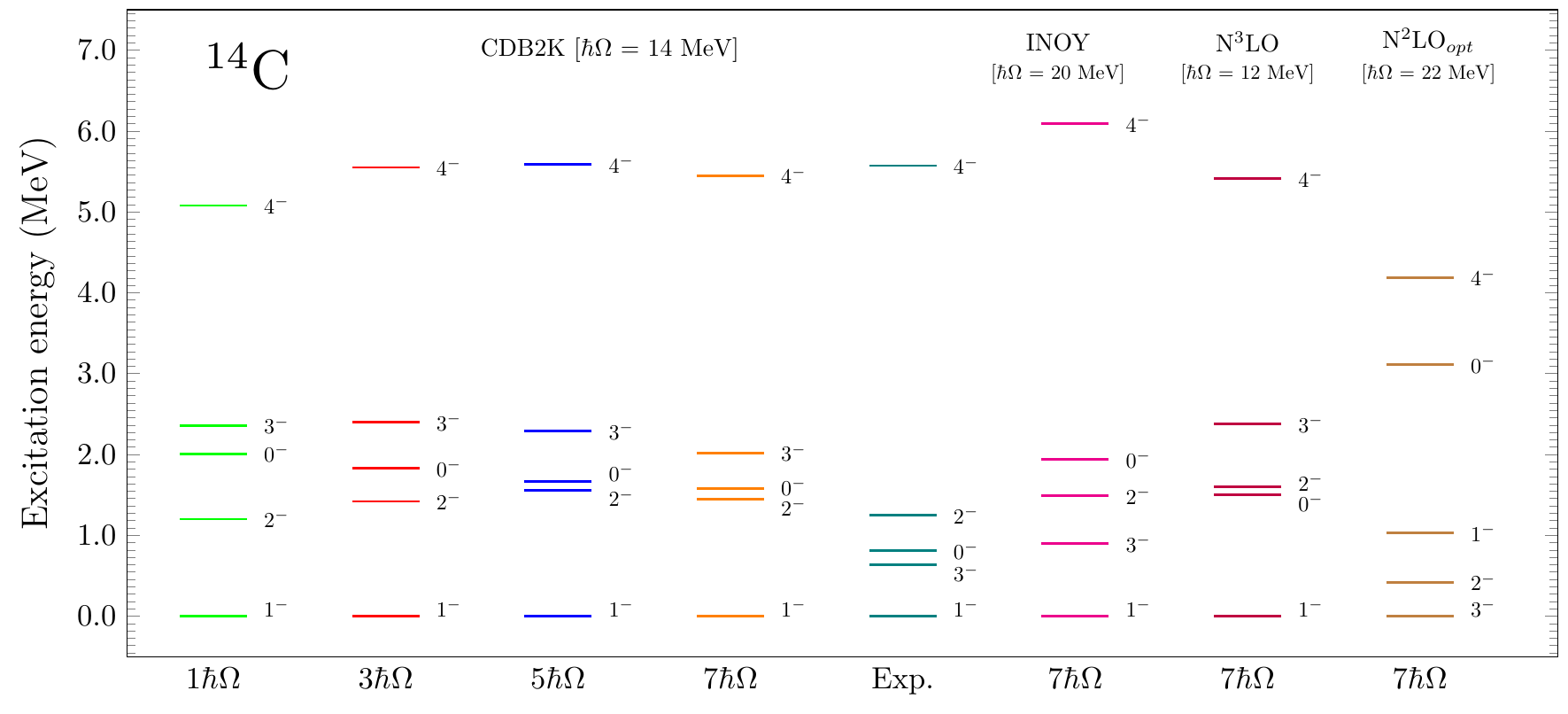}
	\caption{\label{fig:neg_spectra_12_14C}  Excited unnatural parity spectra for $^{12,14}\text{C}$ obtained from the NCSM at their respective optimal frequencies, utilizing four interactions up to $N_\mathrm{max}=7$. Convergence in $N_\mathrm{max}$ is shown using the CDB2K interaction.  Experimental energies are taken from Ref.~\cite{NNDC}.}
\end{figure*}

\begin{figure*}
	\includegraphics[width=17cm]{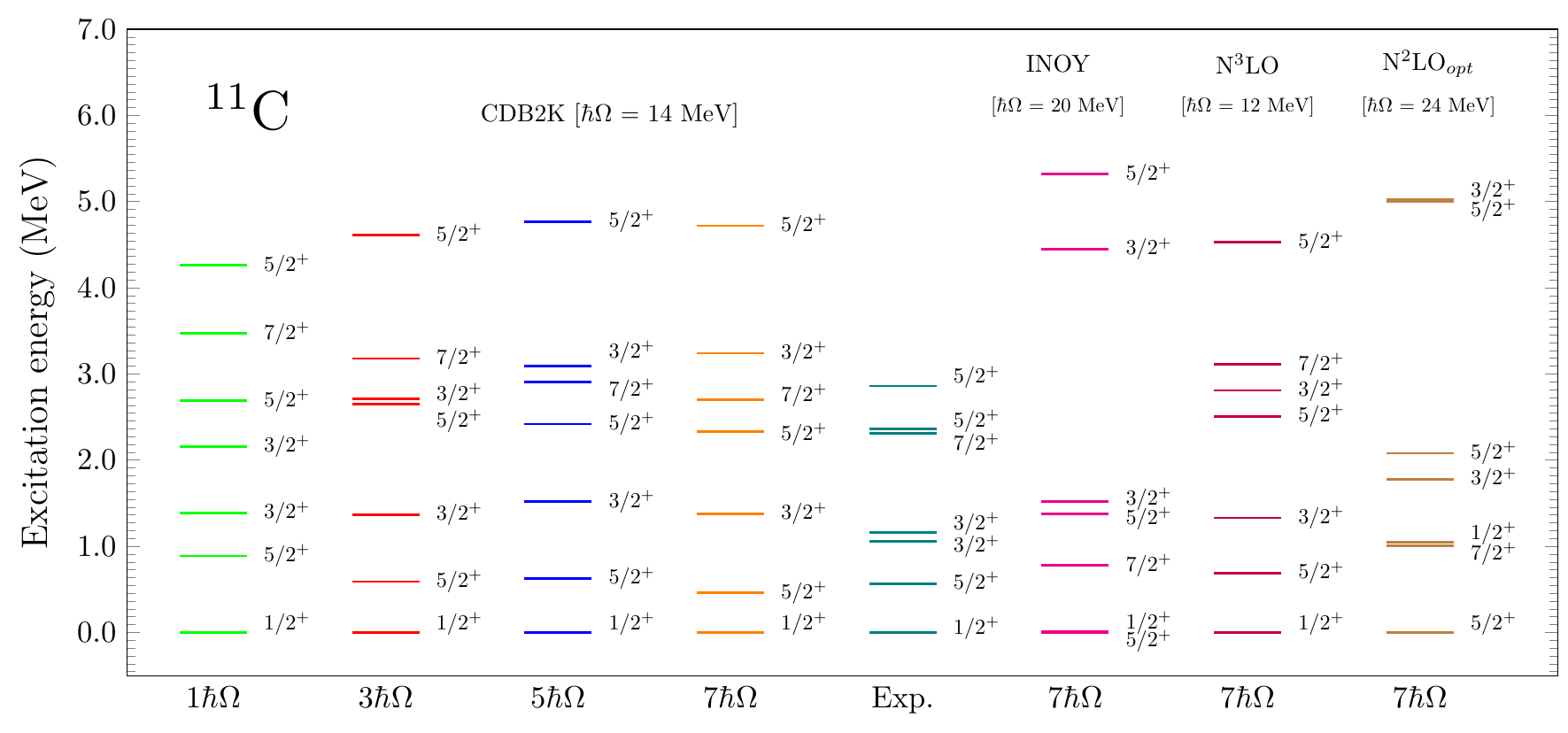}
	\includegraphics[width=17cm]{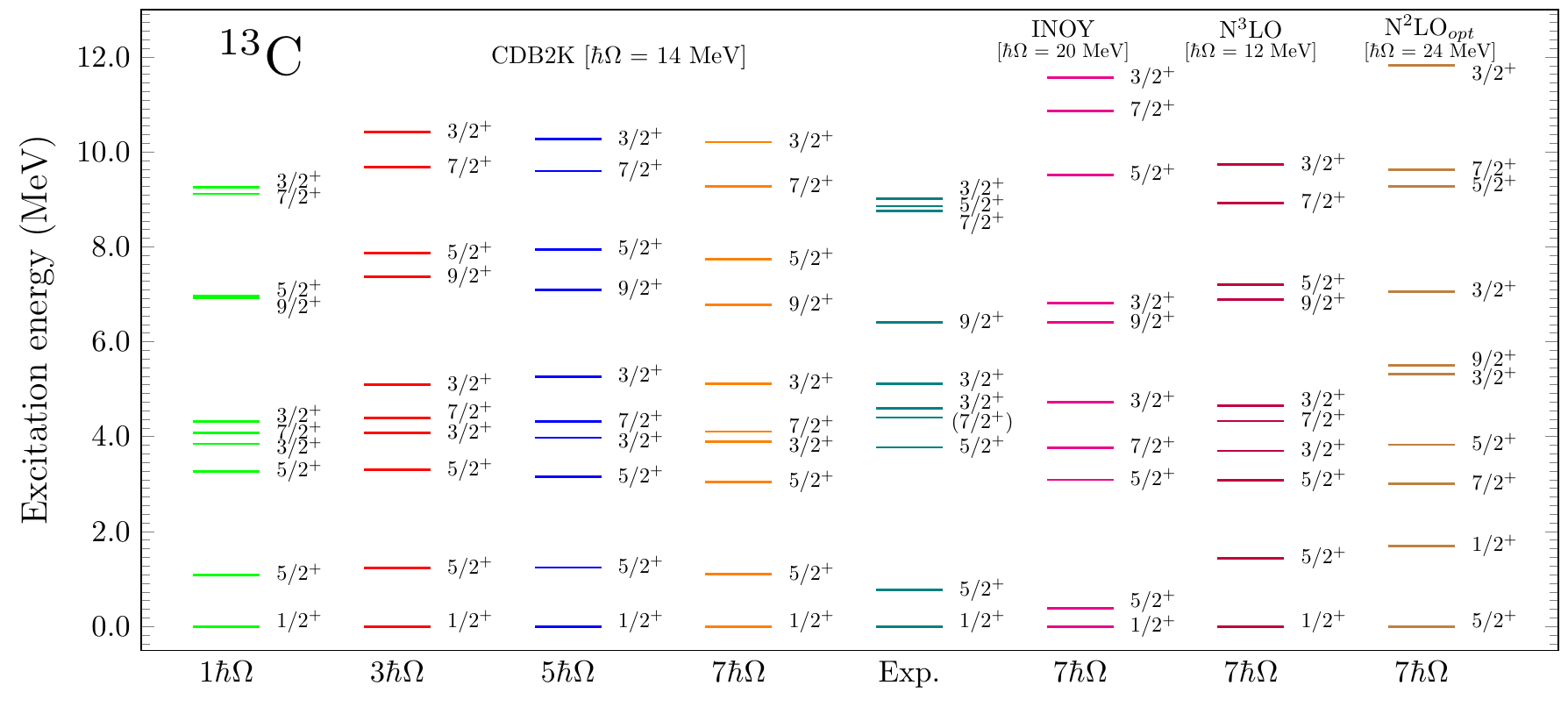}
	\caption{\label{fig:neg_spectra_13C}  Excited unnatural parity spectra for $^{11,13}\text{C}$ obtained from the NCSM at their respective optimal frequencies, utilizing four interactions up to $N_\mathrm{max}=7$. Convergence in $N_\mathrm{max}$ is shown using the CDB2K interaction.  Experimental energies are taken from Ref.~\cite{NNDC}.}
\end{figure*}

In the present work, we  discuss results for the natural  as well as unnatural parity states of the carbon isotopes. These results are presented for $^{10,12,14}\text{C}$ in  Figs.~\ref{fig:spectra_10_12_14C} and \ref{fig:neg_spectra_12_14C} and for $^{11,13}\text{C}$ in  Figs.~\ref{fig:spectra_11_13C} and \ref{fig:neg_spectra_13C}, along with the available experimental spectra. For the CDB2K interaction, we show the convergence of the spectra with $N_\mathrm{max}$ beginning from $N_\mathrm{max}=0$ to the maximal value. For all other interactions, the results are shown  for the largest $N_\mathrm{max}$ basis space.  We indicate the optimal frequencies for each interaction at the top of the figures.
 \vspace{-0.4cm}
\subsection{ Natural parity energy states for $^{10-14}\text{C}$}

\begin{figure*}
\includegraphics[width=17cm]{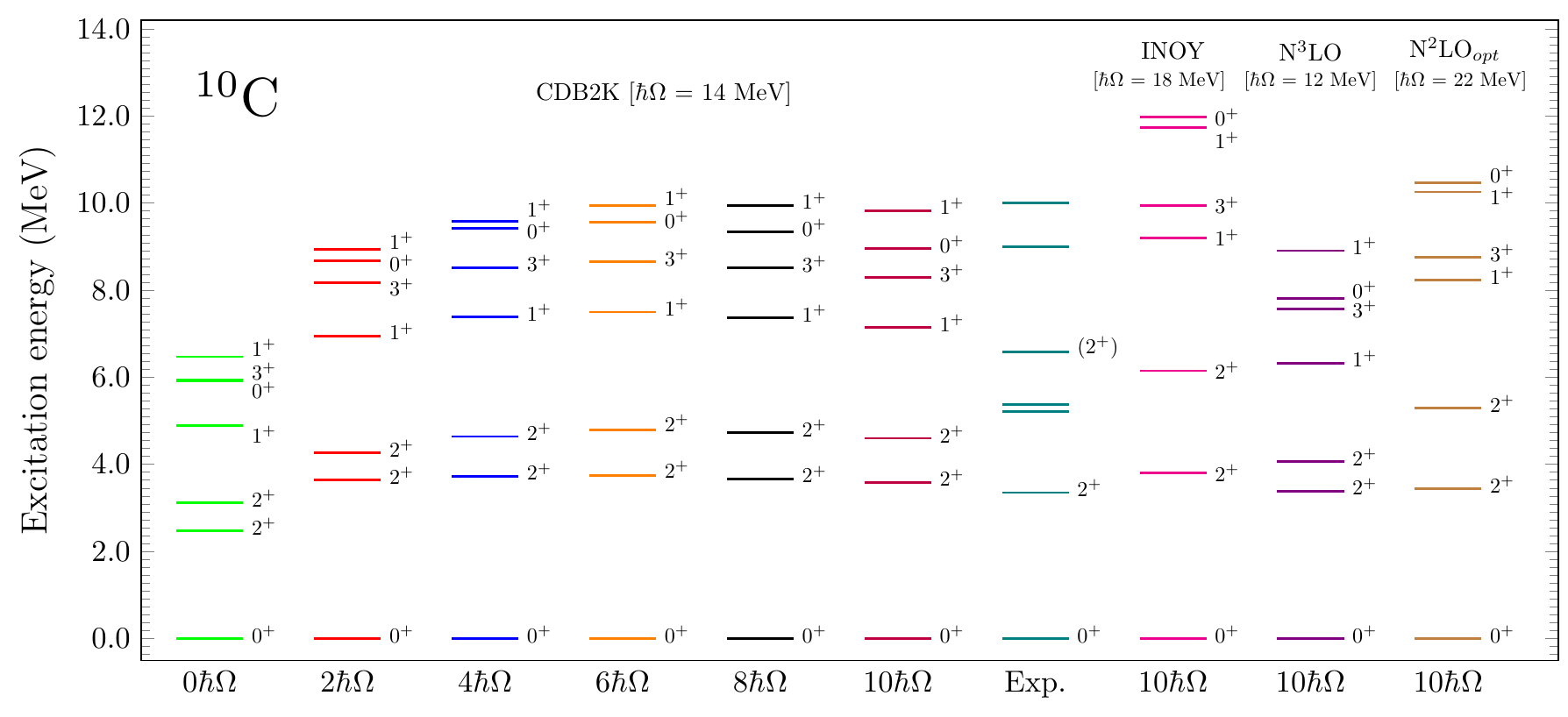}
\includegraphics[width=17cm]{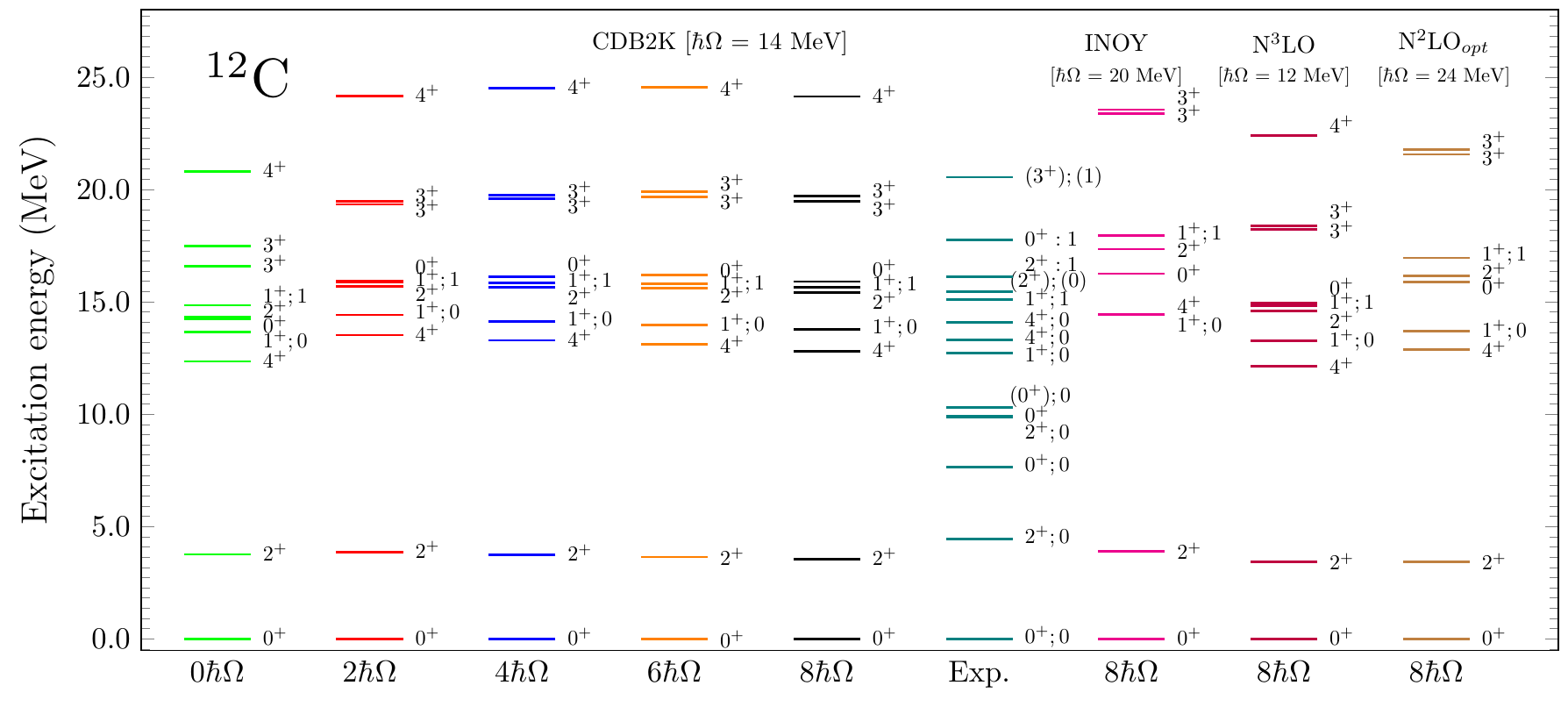}
\includegraphics[width=17cm]{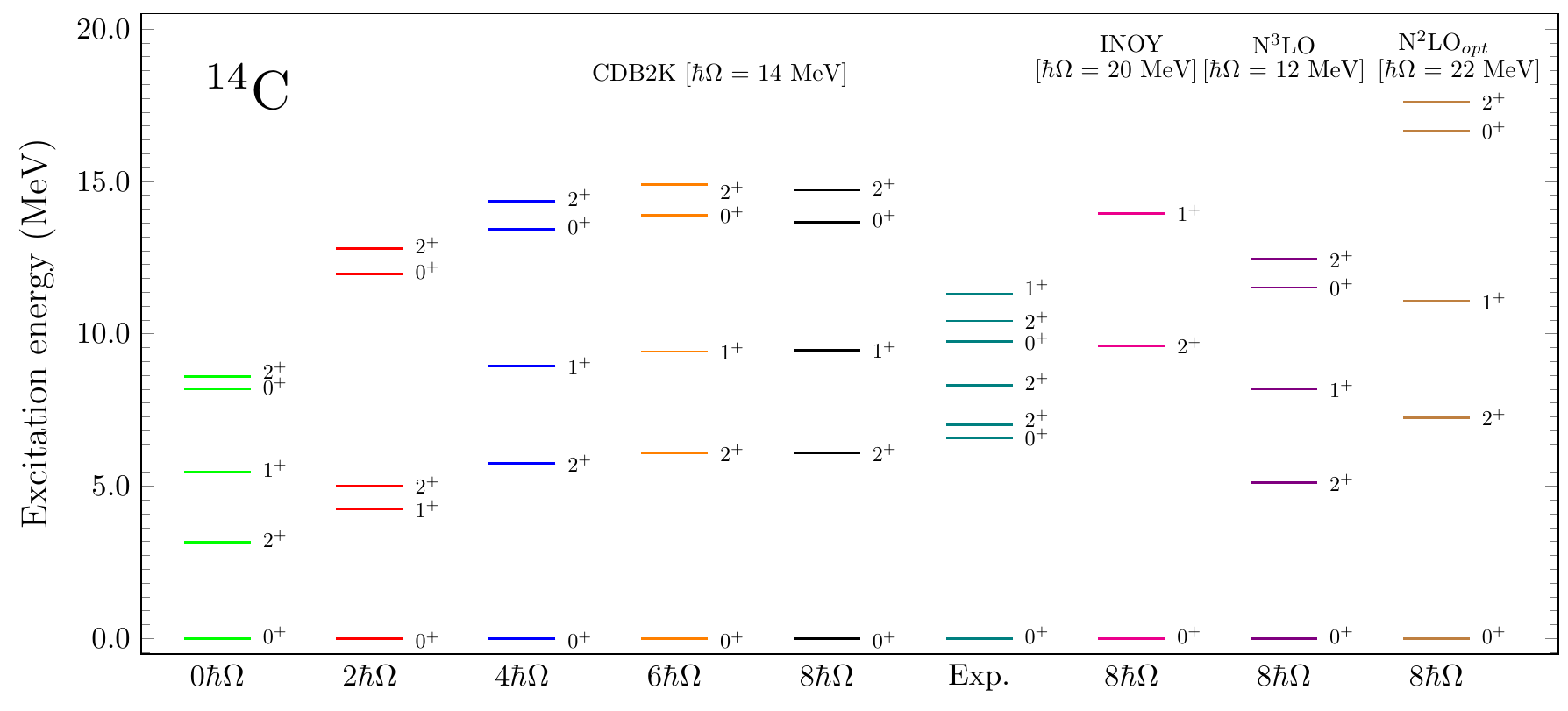}
\caption{\label{fig:spectra_10_12_14C} Excited state spectra for $^{10,12,14}\text{C}$ obtained from the NCSM at their respective optimal frequencies, utilizing four interactions: (i) CDB2K (ii) INOY (iii) N\textsuperscript{3}LO and (iv) N\textsuperscript{2}LO\textsubscript{$opt$}. $^{10}\text{C}$ is calculated at $N_\mathrm{max}=10$, while $^{12,14}\text{C}$ are computed at $N_\mathrm{max}=8$. Convergence in $N_\mathrm{max}$ is shown using the CDB2K interaction.  Experimental energies are taken from Ref.~\cite{NNDC}.}
\end{figure*}

\begin{figure*}[t!]
	\includegraphics[width=17cm]{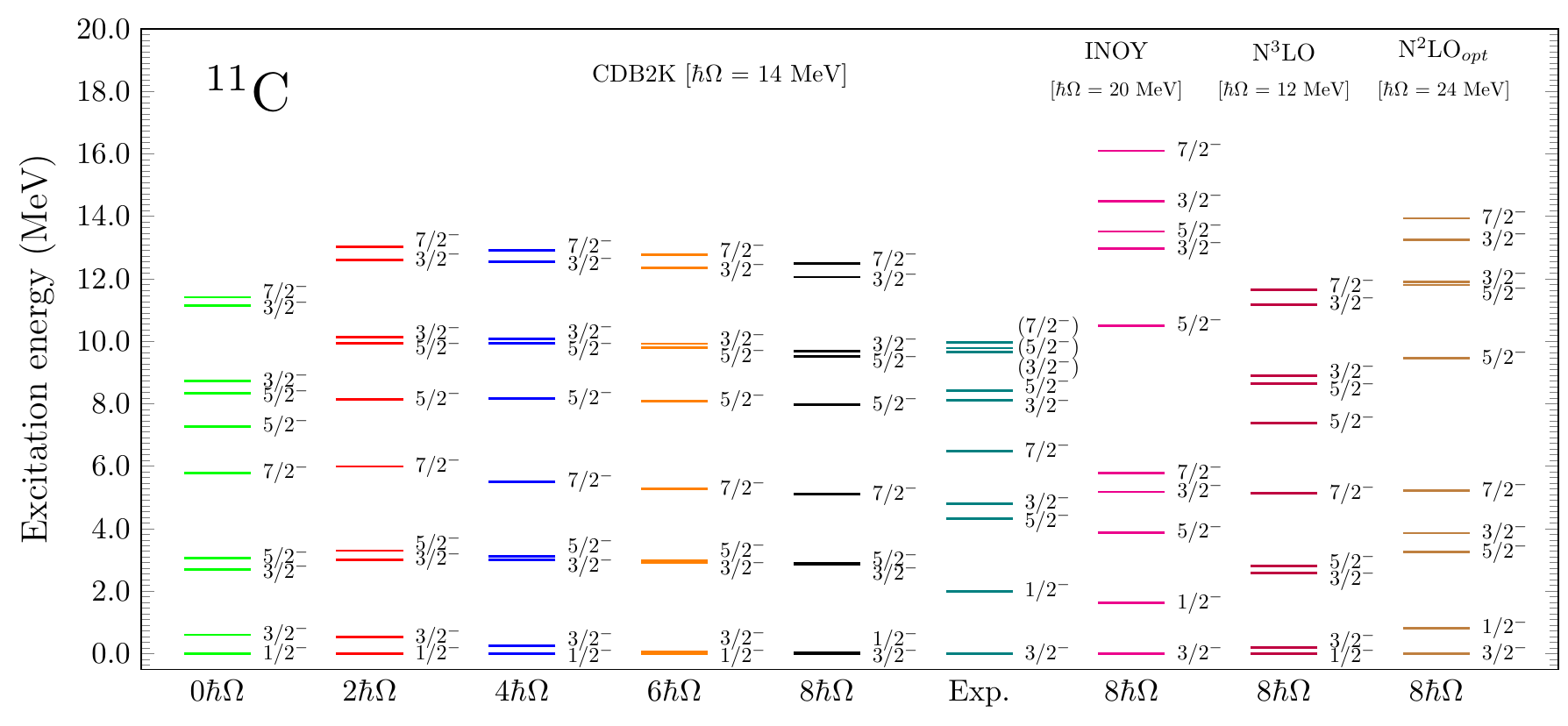}
	\includegraphics[width=17cm]{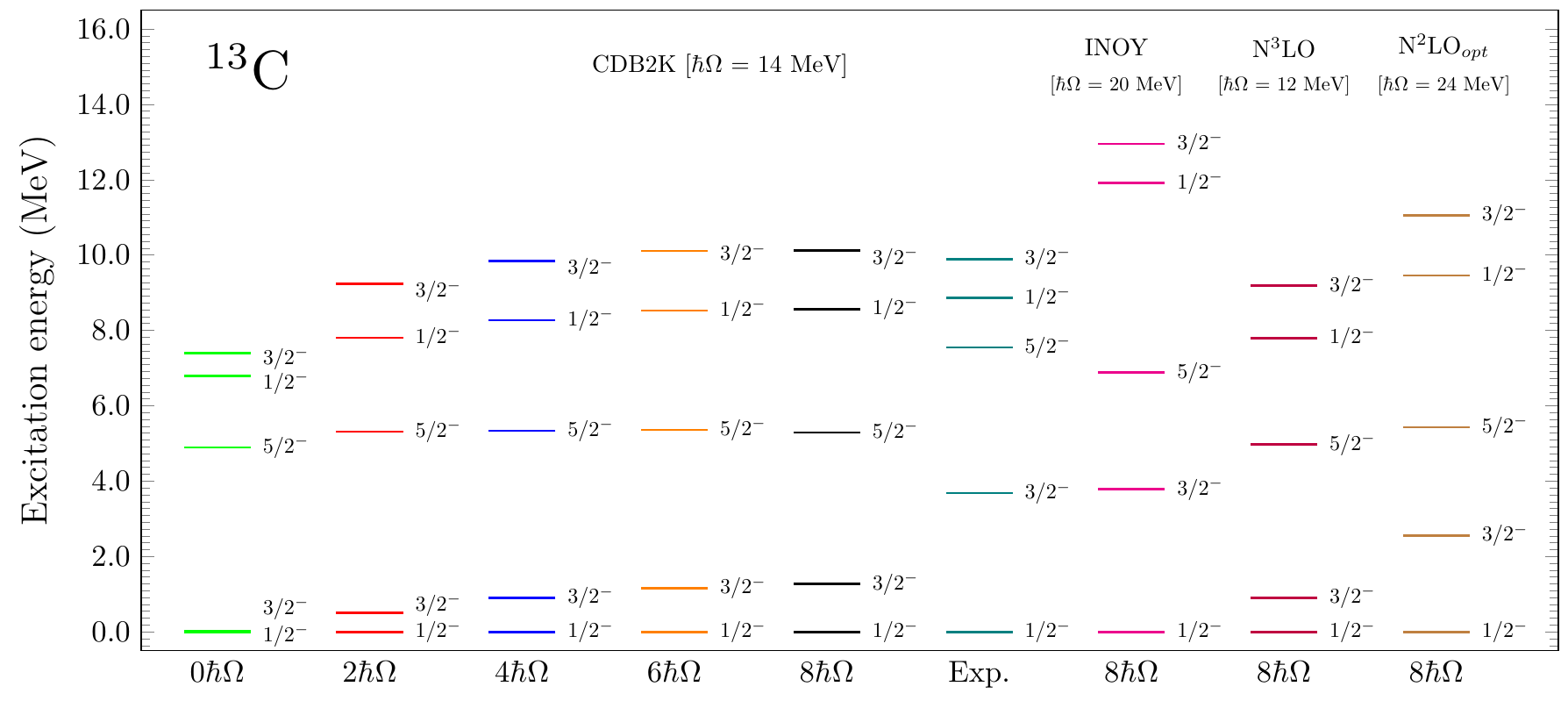}
	\caption{\label{fig:spectra_11_13C} Excited state spectra for $^{11,13}\text{C}$ obtained from the NCSM at their respective optimal frequencies, utilizing four interactions: (i) CDB2K (ii) INOY (iii) N\textsuperscript{3}LO and (iv) N\textsuperscript{2}LO\textsubscript{$opt$}. $^{11,13}\text{C}$ are calculated at $N_\mathrm{max}=8$. Convergence in $N_\mathrm{max}$ is shown using the CDB2K interaction.  Experimental energies are taken from Ref.~\cite{NNDC}.}
\end{figure*}
\begin{figure}
	\includegraphics[width=\columnwidth]{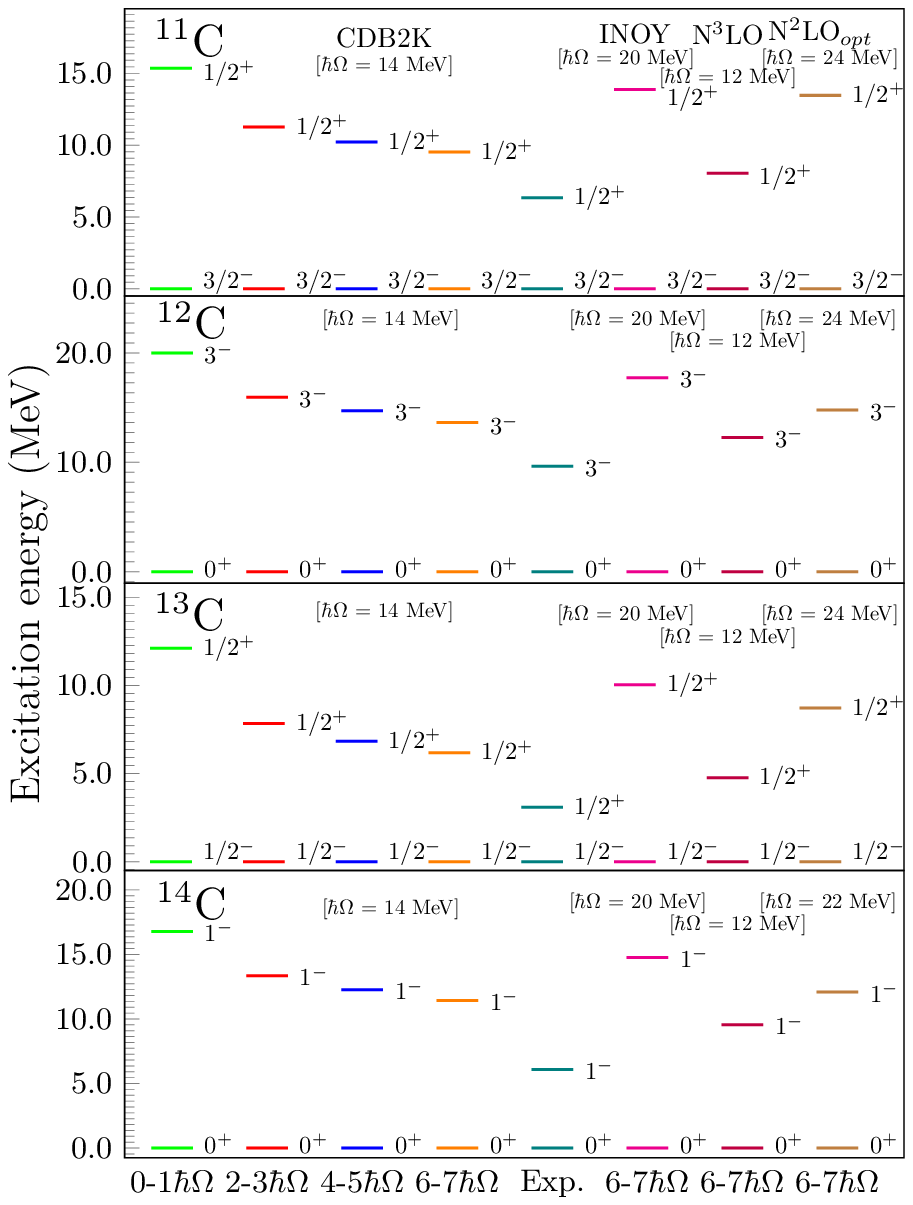}
	\caption{\label{merge}  Excitation energy of lowest unnatural parity states of $^{11,12,13,14}$C relative to the g.s. (lowest natural parity state) in the model space from 0$\hbar$$\Omega$ to 7$\hbar$$\Omega$ for CDB2K. For other interactions, 6$\hbar$$\Omega$ and 7$\hbar$$\Omega$ model space results are shown. Experimental energies are from Ref.~\cite{NNDC}.}
\end{figure}
We begin by discussing the spectra for the even--even nuclei $^{10,12,14}\text{C}$. We first look at $^{10}\text{C}$, the  positive partity spectra for which is shown in the top panel of Fig.~\ref{fig:spectra_10_12_14C}. Experimentally, the spin--parity of only the g. s. $0^+_1$ and first excited $2^+$ states is confirmed (shown in the `Exp.' column). Hence, it is a valuable task for an \textit{ab initio} theory to make predictions for the spin--parity of excited states of $^{10}\text{C}$. As expected, the g.s. spin--parity is correctly reproduced by all four interactions. We have calculated many positive parity states with spin ranging from $J=0-4$, however here we discuss solely the first excited state. The experimental energy of the first excited state, $2^{+}_{1}$, is $3.354$ MeV. Studying the convergence of the energy spectra with $N_\mathrm{max}$ using the CDB2K interaction reveals that as we increase the basis space from $6 \, \hbar \Omega$ to $10 \, \hbar \Omega$, the energy of the excited $2^{+}_{1}$ approaches the experimental value. With $N_\mathrm{max}=10$ model space, the excitation energy of the $2^{+}_1$ is $3.590$ MeV for CDB2K, $3.798$ MeV for INOY, $3.388$ MeV for N\textsuperscript{3}LO and $3.440$ MeV for N\textsuperscript{2}LO\textsubscript{$opt$}.  Interestingly, the description from \textit{ab initio} theory with N\textsuperscript{3}LO is more accurate for this state in this nucleus.

Next, we look at $^{12}\text{C}$, the  positive parity spectra for which is shown in the middle panel of Fig.~\ref{fig:spectra_10_12_14C}. The optimal HO frequencies for $^{12}\text{C}$ are identified as $14 \, \hbar \Omega$, $20 \, \hbar \Omega$, $12 \, \hbar \Omega$ and $24 \, \hbar \Omega$ for the CDB2K, INOY, N\textsuperscript{3}LO and N\textsuperscript{2}LO\textsubscript{$opt$} interactions, respectively. The correct spin--parity of the g.s. is reproduced as anticipated. Looking at the first excited $2^+_1$ state, we see that across all interactions the NCSM has a tendency to overbind this particular state. Further, looking at the excited $2^+_2$ state, the energy difference between the NCSM state and the experimental value is large across all interactions. Using the experimental spectra as reference, we see that the NCSM is incapable of reproducing the energy of the excited $0^+_2$ state, i.e. the Hoyle state, which exists at about $7.65$ MeV. The NCSM tends to underbind this state drastically and estimates energy around $15$ MeV for the various interactions. At the reached basis spaces, the structure of the calculated $0^+_2$ state does not have the $\alpha$--clustering developed as the Hoyle state has.  This means that larger model space is required to obtain converged energy for the state with cluster structure. Lastly, we mention the $1^+_1$ and the $4^+_1$ states, both of which are reasonably described within the NCSM with the INOY interaction predicting their correct order.
In the NCSM, the excitation energy of the $4^+_1$ is $12.789$ MeV for CDB2K, $14.450$ MeV for INOY, $12.144$ MeV for N\textsuperscript{3}LO and $12.888$ MeV for N\textsuperscript{2}LO\textsubscript{$opt$}, while experimentally the value sits at $13.300$ MeV. Generally, compared to the experimental spectra, the NCSM produces an energy levels with a significant gap between the first excited state and the next excited state as the experimental $\alpha$-cluster dominated states are not obtained in the accessible $N_{\rm max}$ spaces.
We have calculated $1^+$ eigenstates corresponding to isospin $T {=} 0$ and $T {= }1$. Experimental  $1_1^+$ state has $T{=} 0$ and $1_2^+$ has $T{ =}1$. We obtain 13.782, 14.446, 13.269 and 13.705 MeV excitation energies for $T {=} 0$ $1_1^+$ state from NCSM calculations corresponding to $N_\mathrm{max}=8$ using CDB2K, INOY, N\textsuperscript{3}LO and N\textsuperscript{2}LO\textsubscript{$opt$} interactions, respectively, while experimentally it is measured at 12.710 MeV.   For $N_\mathrm{max}$ = 8, the CDB2K interaction gives energy of $T {=}1$ $1_2^+$ state at 15.665 MeV. The INOY and N\textsuperscript{2}LO\textsubscript{$opt$} underbind the aforementioned state.
The excitation energy for this state is 14.842 MeV obtained from the N\textsuperscript{3}LO interaction, which is close to the experimental value of 15.110 MeV. 
Thus, we can conclude that $1_2^+$ is better described by N\textsuperscript{3}LO interaction than other interactions.  It is also noted that enlargement in the basis size improves the energy of $1_1^+$ and $1_2^+$ states for the CDB2K interaction.

 We look at $^{14}\text{C}$, the  positive parity spectra for which is shown in the bottom panel of Fig.~\ref{fig:spectra_10_12_14C}. Like other nuclei, the NCSM successfully reproduces the g.s. spin--parity with all interactions. Studying the $N_\mathrm{max}$ trends with the CDB2K interaction, the energies of the first excited $2^+_1$ and $1^+_1$ correctly increase  with increasing $N_\mathrm{max}$ towards the respective experimental values. The excited $2^+_1$ state corresponding to CDB2K interaction is in good agreement with the experimental data. Furthermore, the energies of the excited $0^+_2$ and $2^+_2$ states are substantially different from the experimental values as the experimental states are multi-$\hbar\Omega$ dominated and much large NCSM basis spaces would be required to describe them properly. 

Now, we discuss the energy spectra of even-odd nuclei $^{11,13}$C.
The $^{11}\text{C}$  negative parity spectra are shown in the top panel of Fig.~\ref{fig:spectra_11_13C}. The optimal HO frequencies for $^{11}\text{C}$ are identified as $14 \, \hbar \Omega$, $20 \, \hbar \Omega$, $12 \, \hbar \Omega$ and $24 \, \hbar \Omega$ for the CDB2K, INOY, N\textsuperscript{3}LO and N\textsuperscript{2}LO\textsubscript{$opt$} interactions, respectively. Experimentally, the g.s. spin--parity for $^{11}\text{C}$ is ${3/2}^-_1$, with the first excited state being a ${1/2}^-_1$ lying at $2.000$ MeV. Notably, in the case of the CDB2K interaction, we see that with increasing $N_\mathrm{max}$ the energy difference between the  ${1/2}^-_1$ and ${3/2}^-_1$ states decreases, ultimately reversing the order of the states such that the ${3/2}^-_1$ is then correctly determined to be the g.s. at $N_\mathrm{max}=8$. Nevertheless, the ${1/2}^-_1$ lies extremely close to the g.s., indicating a significant amount of overbinding compared to experiment due to the insufficient strength of the spin-orbit interaction. We see a similar squeezing in the excited ${3/2}^-_2$ and ${5/2}^-_1$ states, which essentially overlap in energy at $N_\mathrm{max}=8$. As with the CDB2K, NCSM calculations with the INOY and N\textsuperscript{2}LO\textsubscript{$opt$} interactions successfully reproduce the spin--parity of the g.s. and the first excited state, though with varying degrees of overbinding in the first excited state still present. Still, the best description is obtained by the INOY interaction for low-lying states. The N\textsuperscript{3}LO interaction is unable to produce the correct g.s. and first excited state ordering. It is feasible that an increase in the model space size would change this ordering, as is seen with the CDB2K spectra.  The INOY and N\textsuperscript{2}LO\textsubscript{$opt$} are also able to reproduce the correct ordering of the first five low--lying states of $^{11}\text{C}$.
\begin{table*}
	\centering
	\caption{The electromagnetic properties of $^{10-14}\text{C}$ isotopes obtained with the NCSM, utilizing four interactions: (i) CDB2K (ii) INOY (iii) N\textsuperscript{3}LO and (iv) N\textsuperscript{2}LO\textsubscript{$opt$} are presented. NCSM calculations are performed at the optimal frequency and largest model space identified in the previous section. The g.s. energy, quadrupole moment, magnetic moment, $B(E2)$ and $B(M1)$ are shown in MeV, barn, $\mu_{N}$, $e^2 \, \text{fm}^4$ and $\mu_{N}^2$, respectively. These are compared with experimental values (where possible). Experimental data are taken from Refs.~\cite{NNDC,Qandmag}.}
	\begin{tabular}{ccMMMMM}
		\hline
		\hline
		\vspace{-2.8mm}\\
		&$^{10}$C & Exp. & CDB2K (14MeV) & INOY (18MeV) & N$^3$LO (12MeV) & N$^2$LO$_{opt}$ (22MeV) \\
		\hline 	
		\vspace{-2.8mm}\\
		&	E$_{g.s.}$($0^{+}$)& -60.320 &-51.685  & -58.697&-50.250 & -49.836  \\
		&	Q($2^{+}$) & NA &-0.017 & -0.026 & -0.010 & -0.031 \\
		&	$\mu$($2^{+}$) & NA & 1.348 & 1.058 & 1.439 &1.038 \\
		&	$B(E2$;$2_{1}^{+}$$\rightarrow$$0_{1}^{+}$) &12.157(1919) & 6.891 & 4.469 &8.987 &  4.861 \\
		&	$B(M1$;$1_{1}^{+}$$\rightarrow$$0_{1}^{+}$) & NA & 0.558 & 0.705&0.501 &0.502 \\
		\vspace{-2.8mm}\\
		\hline
		
		\vspace{-2.8mm}\\
		& $^{11}$C& Exp. & CDB2K (14MeV) & INOY (20MeV) & N$^3$LO (12MeV) & N$^2$LO$_{opt}$ (24MeV) \\
		\hline 
		\vspace{-2.8mm}\\
		&	E$_{g.s.}$($3/2^{-}$)& -73.441 & -63.871 &-71.931 &-62.941 &-57.084 \\
		&	Q$_{g.s.}$($3/2^{-}$) & 0.0333(2) & 0.020& 0.022 & 0.018  &0.023  \\
		&	$\mu$$_{g.s.}$($3/2^{-}$) & -0.964(1) & -0.863 &-0.638  &-0.995  &-0.634  \\
		&	$\mu$($1/2^{-}_1$) & NA & 0.905 & 0.890  &0.904  &0.861  \\
		&	$B(E2$;$7/2_{1}^{-}$$\rightarrow$$3/2_{1}^{-}$) &NA & 5.980 &3.940 &7.080 &4.487   \\
		&	$B(M1$;$3/2_{1}^{-}$$\rightarrow$$1/2_{1}^{-}$) & 0.340(27) & 0.754 &0.458 &0.867 &0.505\\
		&	$B(M1$;$5/2_{1}^{-}$$\rightarrow$$3/2_{1}^{-}$) & NA & 0.204 &0.352 &0.138 &0.304 \\
		\vspace{-2.8mm}\\
		\hline

		\vspace{-2.8mm}\\
		& $^{12}$C& Exp. & CDB2K (14MeV) & INOY (20MeV)& N$^3$LO (12MeV)& N$^2$LO$_{opt}$ (24MeV) \\
		\hline 
		\vspace{-2.8mm}\\
		&	E$_{g.s.}$($0^{+}$)&-92.162 &-83.164  & -93.473& -81.725& -75.276  \\
		&	Q($2^{+}$) &0.06(3)  &0.051 & 0.040 & 0.056 & 0.044  \\
		&	$\mu$($2^{+}$) & NA & 1.026 & 1.039 &1.024  &1.034 \\
		&	$B(E2$;$2_{1}^{+}$$\rightarrow$$0_{1}^{+}$) &7.59(42) & 5.942 & 3.644 &7.277 &  4.354\\
		&	$B(M1$;$1_{1}^{+}$$\rightarrow$$0_{1}^{+}$) & 0.0145(21) & 0.005 &0.010 & 0.005& 0.007\\
		\vspace{-2.8mm}\\
		\hline
		
		\vspace{-2.8mm}\\
		& $^{13}$C & Exp. & CDB2K (14MeV) & INOY (20MeV)& N$^3$LO (12MeV)& N$^2$LO$_{opt}$ (24MeV) \\
		\hline 	
		\vspace{-2.8mm}\\
		&	E$_{g.s.}(1/2^-)$& -97.108& -87.308&-99.831&-85.905&-78.058\\
		&	$\mu$$_{g.s.}$($1/2^{-}$) & 0.702369(4) & 0.881 & 0.691 & 0.933 &0.738 \\
		&	Q($3/2^{-}_1$) & NA & 0.032 &0.023 & 0.036 & 0.026\\
		&	$\mu$($3/2^{-}_1$) &NA  & -0.874 & -0.674 & -0.934 &-0.737  \\
		&	$B(E2$;$5/2_{1}^{-}$$\rightarrow$$1/2_{1}^{-}$)  &5.629(363) & 4.812&2.959 & 5.824 & 3.544 \\
		&	$B(M1$;$3/2_{1}^{-}$$\rightarrow$$1/2_{1}^{-}$) &0.698(72) & 1.174&0.872 &1.257 &0.949\\
		\vspace{-2.8mm}\\
		\hline
		\vspace{-2.8mm}\\
		& $^{14}$C & Exp. & CDB2K (14MeV) & INOY (20MeV) & N$^3$LO (12MeV) & N$^2$LO$_{opt}$ (22MeV) \\
		\hline
		\vspace{-2.8mm}\\
		&	E$_{g.s.}$($0^{+}$)& -105.284&-97.753  & -111.239& -96.129& -86.229  \\
		&	Q($2^{+}$) & NA & 0.042 & 0.031 & 0.048 & 0.037 \\
		&	$\mu$($2^{+}$)& NA &2.397  &2.550  &2.346  &2.454 \\
		&	$B(E2$;$2_{1}^{+}$$\rightarrow$$0_{1}^{+}$) &3.608(602) &  4.396&2.673  &5.473 &3.529 \\
		&	$B(M1$;$1_{1}^{+}$$\rightarrow$$0_{1}^{+}$) & 0.394(89) &0.9167  & 1.215&0.792 &1.078  \\
		\vspace{-2.8mm}\\
		\hline
		
		\hline
		\hline
	\end{tabular}
	\label{tbl:EM}
\end{table*}

Next, we look at $^{13}\text{C}$, the negative parity spectra for which is shown in the bottom panel of Fig.~\ref{fig:spectra_11_13C}. The optimal HO frequencies are exactly the same as in the case of $^{11}\text{C}$. Experimentally  the g.s. spin--parity is  ${1/2}^-;T=1/2$, which is  reproduced by all interactions. Interestingly, looking at the CDB2K spectra convergence trends, we see that the g.s. ${1/2}^-_1$ and excited ${3/2}^-_1$ are nearly degenerate at $N_\mathrm{max}=0$. However, by $N_\mathrm{max}=8$ the ${3/2}^-_1$ has moved substantially far from the g.s. in the correct direction, yet it does not shift nearly close enough to the true experimental value. 
The INOY interaction on the other hand reproduces the g.s. and first excited state quite well, with the NCSM energy for the ${3/2}^-_1$ being $3.792$ MeV compared to the experimental value of $3.684$ MeV. The low--lying states are better described using the INOY interaction, however the remaining spectra is not as dense as expected.

\subsection{ Unnatural parity energy states of $^{11-14}$C} 

In this section, we present the energy spectra of unnatural parity states, which are negative parity states in the case of even-even carbon isotopes and positive for even-odd isotopes. Negative parity energy spectra for $^{12}$C and $^{14}$C are shown in Fig. \ref{fig:neg_spectra_12_14C} and positive parity spectra for $^{11}$C and $^{13}$C in fig. \ref{fig:neg_spectra_13C}, calculated at same optimal frequency as for the natural parity states. Experimental negative parity states are unknown for $^{10}$C. Thus, we have not performed calculations for $^{10}$C negative parity states. The ordering of low-lying negative parity states up to $4_1^-$ for $^{12}$C is obtained in the correct sequence similar to the experiment using CDB2K, INOY, and N$^3$LO interactions. The energy difference between the lowest $T=0$ $3^-$ and $1^-$ states is higher for all interactions compared to the experimental difference, especially in the case of bare N$^2$LO$_{opt}$ interaction. Similarly, 
$2^-$ and $4^-$ states are obtained at high excitation energies. 
Experimentally, the first lowest negative parity state for $^{14}$C is $1^-$ that is reproduced by all interactions except for N$^2$LO$_{opt}$. Results of excited negative parity states obtained with CDB2K are improved with increasing basis size. 
In the case of $^{11}$C, CDB2K interaction gives the lowest positive parity state as $1/2^+$, which is in agreement with the experimental spin. The $5/2^+$ state is obtained as the lowest positive parity state with INOY and N$^2$LO$_{opt}$  interactions. With INOY interaction, the energy difference between $5/2^+$ and $1/2^+$ states is 6.6 keV at $N_\mathrm{max} = 7$. Since this difference is very small and decreased with increasing $N_\mathrm{max}$, one could expect to get the correct order of these states as in the experiment for a larger basis size.     
For CDB2K, INOY, and N$^3$LO interactions, the first three lowest positive parity states are in the same order as for the experiment for $^{13}$C. The order of $^{13}$C $5/2^+$ and $1/2^+$ states is reversed using N$^2$LO$_{opt}$ compared to the experimental result. The first $(7/2^+)$ state at 7.492 MeV is confirmed from our theoretical NCSM calculations. The energy splitting between $9/2_1^+$ and $5/2_3^+$ states is 2.45 MeV, experimentally. We can see from the spectrum that, at $N_\mathrm{max} = 1$, these two states are almost degenerate. This energy splitting between these two states starts to increase as we move to the higher basis size using CDB2K interaction, and it becomes around 1 MeV at $N_\mathrm{max} = 7$.
		
In Fig. \ref{merge}, the lowest both positive and negative parity states of $^{11-14}$C are presented in the model space from 0$\hbar\Omega$ to 7$\hbar\Omega$ in the case of the CDB2K, and results corresponding to 6$\hbar\Omega$ and 7$\hbar\Omega$ model spaces are shown for other three interactions. The excitation energy of unnatural parity states in $N_\mathrm{max}\hbar\Omega$ model space is taken relative to the natural parity g.s. in $(N_\mathrm{max}-1)\hbar\Omega$ space. We found from the convergence of spectra for CDB2K interaction that, with basis size enlargement, excitation energies of unnatural parity states improve for all carbon isotopes. For further improvement, the basis size needs to be extended. Excitation energy obtained using INOY interaction for unnatural parity state is quite large in comparison with the experimental excitation energy.

\section{Electromagnetic Properties}\label{sec:em_results}
\begin{figure}[t]
\includegraphics[width=8.5cm]{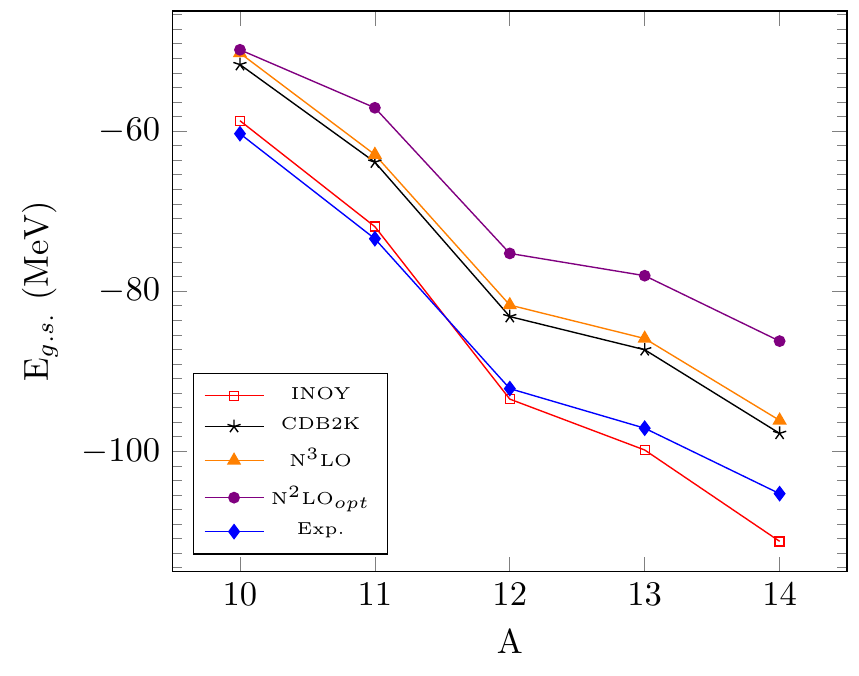}
\caption{\label{fig:Egs_vs_A} The g.s. energies for $^{10-14}\text{C}$ obtained from the NCSM as a function of mass number $A$, utilizing four interactions: (i) CDB2K (ii) INOY (iii) N\textsuperscript{3}LO and (iv) N\textsuperscript{2}LO\textsubscript{$opt$}. NCSM calculations are performed at the optimal frequency and largest model space identified in the previous section. These are compared with experimental data \cite{NNDC}.}
\end{figure}

In Table~\ref{tbl:EM}, we report g.s. energies, quadrupole moments, magnetic moments, reduced electric quadrupole transition strengths and reduced magnetic dipole transition strengths of the carbon isotopes obtained from the \textit{ab initio} NCSM calculations using the four NN interactions. These properties were obtained using the optimal oscillator frequencies and largest possible $N_\mathrm{max}$ basis space. We provide the corresponding experimental data for comparison.

We first discuss the g.s. energies briefly, using $^{10}\text{C}$ as an example. The energy of the g.s. of $^{10}\text{C}$ is $-60.320$ MeV according to experiment. As shown in Table~\ref{tbl:EM}, the NCSM calculations with the various interactions yield the following results: $-51.685$ MeV for CDB2K, $-58.697$ MeV for INOY, $-50.250$ MeV for N\textsuperscript{3}LO and $-49.836$ MeV for N\textsuperscript{2}LO\textsubscript{$opt$}.  This data indicates that the INOY result is close to the experimental value. 
Referring to Fig.~\ref{fig:Egs_vs_A}, we have plotted the g.s. energy of the carbon isotopes against mass number for all aforementioned interactions. The plot reinforces the fact that within the set of \textit{ab initio} calculations, use of the INOY interaction provides the best g.s. energy agreement with the experiment.

\begin{figure*}
	\includegraphics[width=8.75cm]{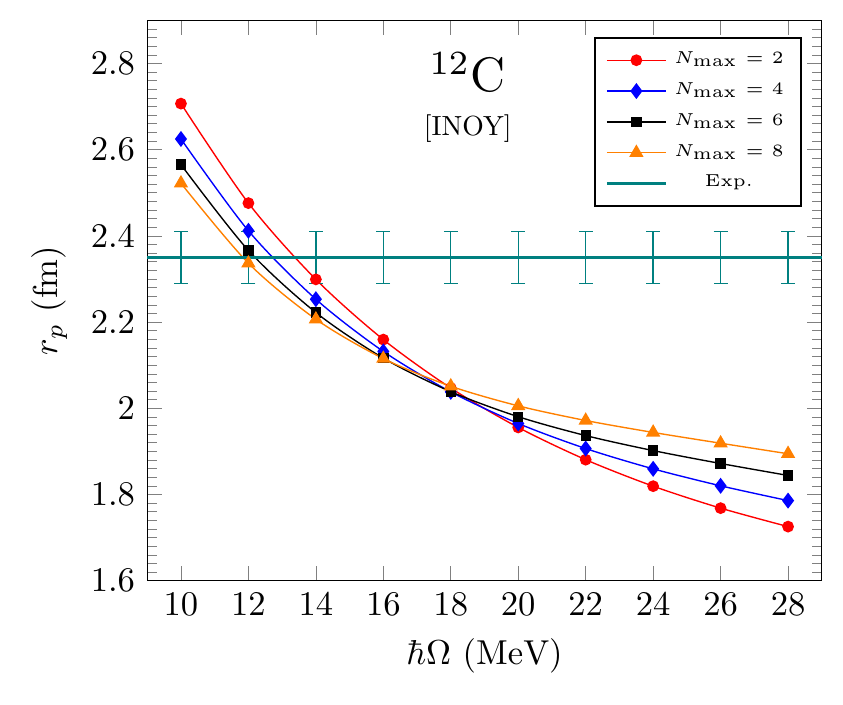}
	\includegraphics[width=8.75cm]{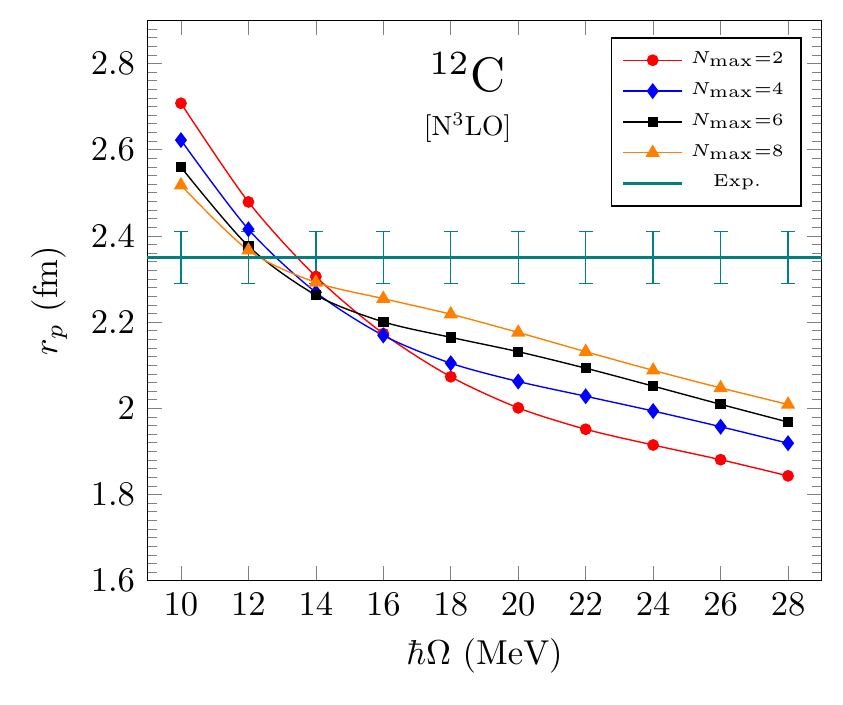}
	\caption{\label{fig:pradii_INOY} Point--proton radii computed within the NCSM utilizing the INOY and N\textsuperscript{3}LO interactions for $^{12}\text{C}$, using a range of oscillator frequencies from 10--28 MeV. The model space is varied from $N_\mathrm{max}=2-8$. Experimental radii are taken from Ref.~\cite{RCNP} and shown as a horizontal line with error bars.}
\end{figure*}

For the even $A$ nuclei, we have performed calculations of the quadrupole and magnetic moments for the first excited state of each nucleus (for the even carbon isotopes discussed, this is a $2^+_1$ state). Experimental data for the  $Q(2^+_1)$ exists only for $^{12}\text{C}$. The sign of $Q$ is correct for each interaction, however with the NCSM the magnitude is best reproduced by the CDB2K and the N\textsuperscript{3}LO interactions. No experimental data for the $\mu(2^+_1)$ of $^{10,12,14}\text{C}$ are available. The calculations of $\mu(2^+_1)$ are substantially consistent with one another. We also calculate the reduced electric quadrupole transition strength $B(E2;\, 2^+_1 \rightarrow 0^+_1)$ for $^{10,12,14}\text{C}$ and compare them to experiment. For $^{10}\text{C}$, the $B(E2)$ value is far from experiment regardless of the interaction chosen. 
The $B(E2)$ result is in good agreement with the experimental data for N\textsuperscript{3}LO interaction in $^{12}\text{C}$, while for N\textsuperscript{2}LO\textsubscript{$opt$} interaction in the case of $^{14}\text{C}$.
 The $B(E2)$ transition strength from the first excited state $2_1^+$ to the g.s. $0^+$ is reduced from $^{10}$C to $^{14}$C. As the number of neutrons is increased from $N=4$ to $N=8$, experiment suggests a decrease in collectivity. This trend is similarly seen with each interaction. The strength of the $B(E2)$ transitions with the \textit{ab initio} interactions is somewhat different from the experimental data. This is due to the fact that the $B(E2)$ is a long--range operator and the OLS unitary transformation only renormalizes the short range part of the interaction and short--range operators, while long--range operators are weakly renormalized. Thus, larger model space sizes are required to obtain converged results for the $B(E2)$ transition.  Another way to obtain convergence of $E2$ observables is mentioned in Refs. \cite{Mark1,Mark2}.

Similarly we have calculated the reduced magnetic dipole transition strength $B(M1;\, 1^+_1 \rightarrow 0^+_1)$ for the even nuclei. Experimental data exists only for $^{12,14}\text{C}$, thus, the $B(M1;\, 1^+_1 \rightarrow 0^+_1)$ transitions are predicted in the NCSM for $^{10}\text{C}$. In $^{12}\text{C}$, the situation is under control with calculations providing a reasonable degree of agreement with one another and with the experimental value. However, in the case of $^{14}\text{C}$ we see a significant variance in the calculations, all of which drastically overpredict the transition strength. 

For $^{11}\text{C}$, we have performed calculation of the quadrupole moment of the g.s., and the magnetic moments of the g.s. and first excited state. For the g.s. quadrupole moment $Q({3/2}^-_1)$, we find that the NCSM calculations are generally consistent with one another across all interactions. Looking at the magnetic moment $\mu_{g.s.}({3/2}^-_1)$, we find that the sign is correctly reproduced for each interaction, however, the NCSM results with the CDB2K and N\textsuperscript{3}LO interactions are close to the experimental value.   
The NCSM result of $B(M1;\, {3/2}^-_1 \rightarrow {1/2}^-_1)$ corresponding to INOY and 
N\textsuperscript{2}LO\textsubscript{$opt$} are in a reasonable agreement with the experimental data. 
It is worth noting that $B(M1;\, {5/2}^-_1 \rightarrow {3/2}^-_1)$ transition strength is predicted in the NCSM for $^{11}\text{C}$ and have yet to be experimentally observed. Moving on to $^{13}\text{C}$, we have computed the magnetic moment of the g.s. and first excited state, as well as the quadrupole moment for the first excited state. For the g.s. magnetic moment $\mu_{g.s.}({1/2}^-_1)$, we see the NCSM description are in good agreement with the experimental data. We have also computed the $B(E2;\, {5/2}^-_1 \rightarrow {1/2}^-_1)$ and $B(M1;\, {3/2}^-_1 \rightarrow {1/2}^-_1)$ transition strengths.

\section{Point--proton radii}\label{sec:protonradii_results}
In addition, we have investigated the point--proton radii $r_p$ for the carbon isotopes, the results of which are tabulated (using the optimal frequency and largest $N_\mathrm{max}$ model space) in Table~\ref{tbl:radii}. The results are shown alongside the experimental values~\cite{RCNP}. Comparing to experiment where possible, it seems that the INOY interaction is incapable of producing  radii. On the other hand, the CDB2K and N\textsuperscript{3}LO interactions produce more realistic results, with the CDB2K radii being smaller by about $4\%$ and the N\textsuperscript{3}LO giving more accurate description with a deviation from experiment ranging from $0.4\%$ to $1.8\%$.
\vspace{-0.5cm}
\begin{table}[!h]
	\centering
	\caption{\label{tbl:radii} Point--proton radii for $^{10-14}\text{C}$ obtained from the NCSM, utilizing three interactions: (i) CDB2K (ii) INOY and (iii) N\textsuperscript{3}LO. The point--proton radii are shown in fm. NCSM calculations are performed at the optimal frequency and largest model space identified in the previous section. Experimental data are taken from Ref.~\cite{RCNP}.}
	\begin{tabular}{cMMMM}
		\hline
		\hline \vspace{-2.8mm}\\
		$ r_{p}$ & Exp. & CDB2K & INOY & N$^3$LO \\
		\hline \vspace{-2.8mm}\\
		$^{10}$C & NA &2.38 & 2.18 &2.50  \\
		$^{11}$C & NA &2.29 &2.04  &2.41   \\
		$^{12}$C & 2.35(6) &2.25  & 2.01& 2.37 \\
		$^{13}$C & 2.33(13) &2.23  &1.96 & 2.34 \\
		$^{14}$C & 2.27(6) &2.19 &1.94  &2.31   \\
		\hline \hline
	\end{tabular}
\end{table}

It is generally quite challenging for \textit{ab initio} methods to reliably predict the point--proton radii of nuclei. The radii are highly sensitive to the interaction between nucleons and to long--range features of the wave function.  In Fig.~\ref{fig:pradii_INOY}, we present NCSM calculations of $r_p$ for $^{12}\text{C}$ using the INOY and N\textsuperscript{3}LO interactions, ranging the oscillator frequency from 10--28 MeV and the model spaces from $N_\mathrm{max}=2-8$. Different curves correspond to different $N_\mathrm{max}$ calculations and the experimental values are shown as solid horizontal lines.   As can be seen, the NCSM calculations of $r_p$ are extremely sensitive to the NCSM parameters. As basis size increases from 2$\hbar$$\Omega$ to 8$\hbar$$\Omega$, decrease of $r_p$ dependence on HO parameters is observed. Arguably, the best results are obtained with the N\textsuperscript{3}LO interaction which tends to produce the flattest curve at $N_\mathrm{max}=8$ even compared to CDB2K calculations (not shown), indicating a greater independence with respect to the HO frequency.

 Radii curves with different $N_\mathrm{max}$ basis spaces cross one another at approximately the same frequency. As suggested in Ref.~\cite{Radii1,Radii2}, this crossing point is estimated to be the converged point--proton radii. In our case, we consider the intersection point of largest two $N_\mathrm{max}$ curves as the converged proton radii. From Fig.~\ref{fig:pradii_INOY}, the converged point--proton radii for $^{12}$C is obtained as 2.11 fm and 2.34 fm for INOY and N\textsuperscript{3}LO interactions, respectively. We see that the N\textsuperscript{3}LO interaction predicts the radii of $^{12}$C to be quite close to experiment. Similarly, $r_p$ calcuated with N$^3$LO for the other isotopes is reasonably consistent with experiment. We also observe that optimal frequency corresponding to converged point--proton radii is smaller than that obtained from the g.s. energy minima.

\vspace*{-0.7cm}
\section{NO-CORE SHELL MODEL densities}\label{sec:ncsm_densities}
\vspace*{-0.4cm}
\begin{figure*}
 	\includegraphics[width=8.75cm]{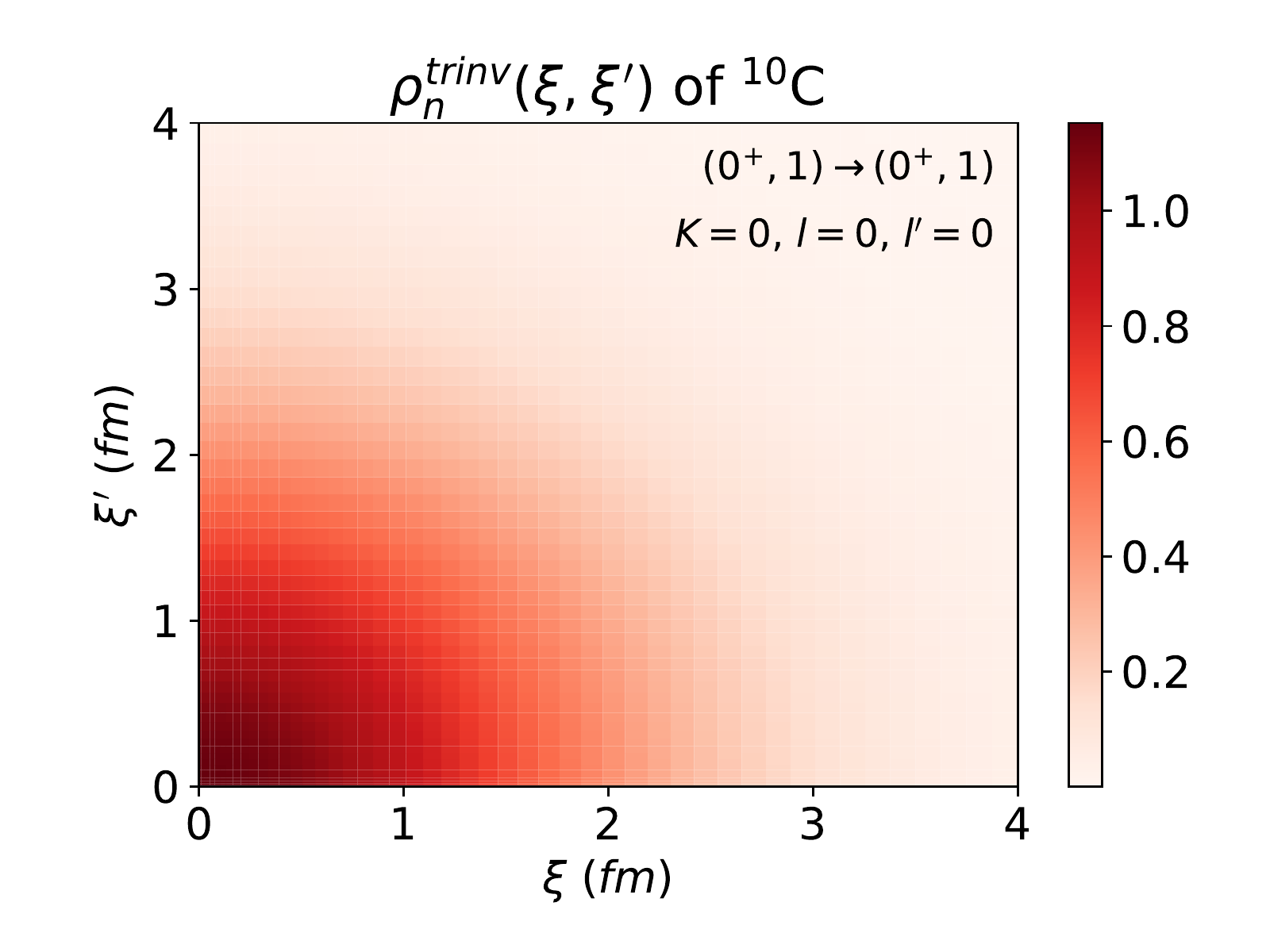}
 	\includegraphics[width=8.75cm]{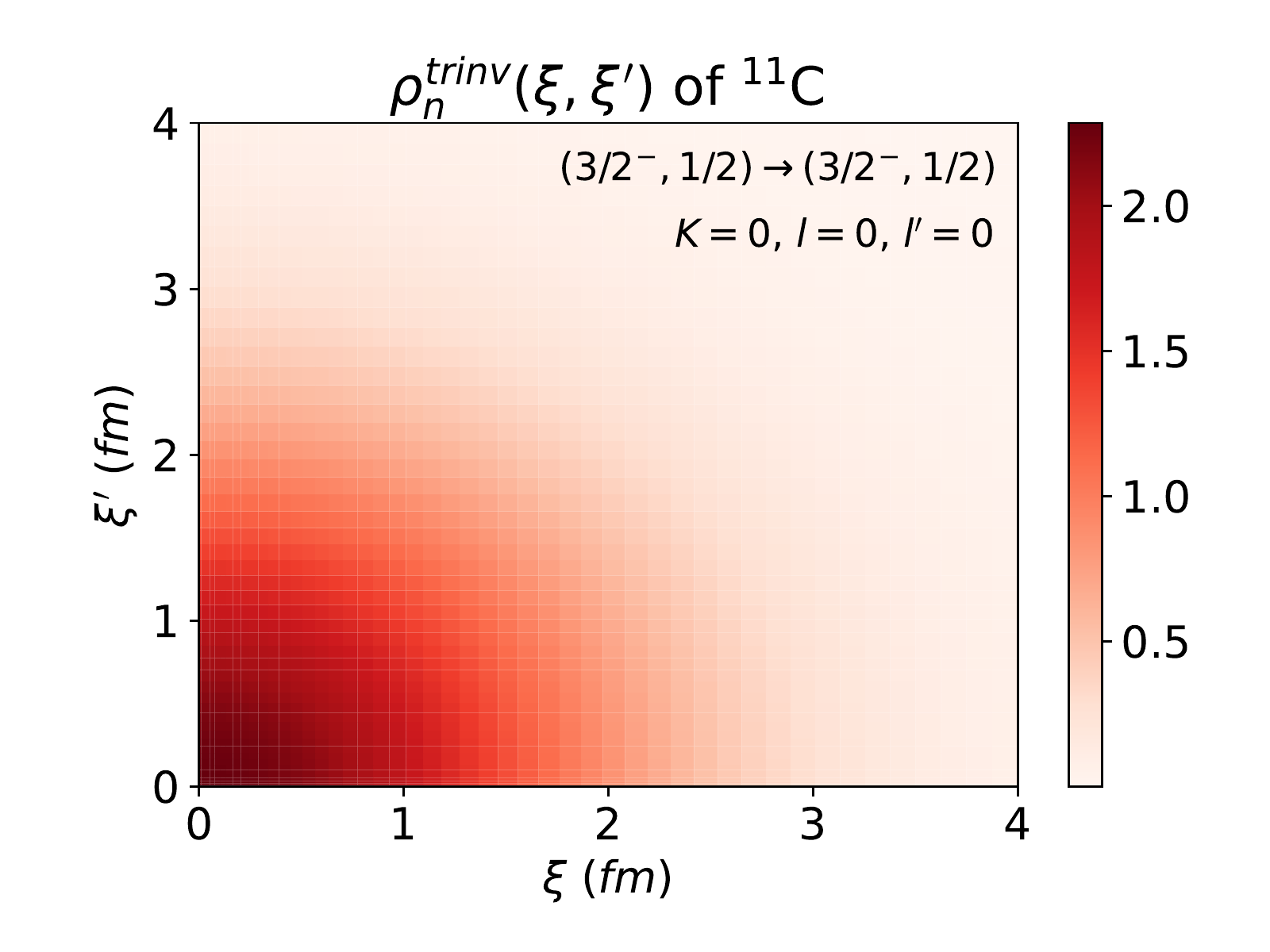}
 	\includegraphics[width=8.75cm]{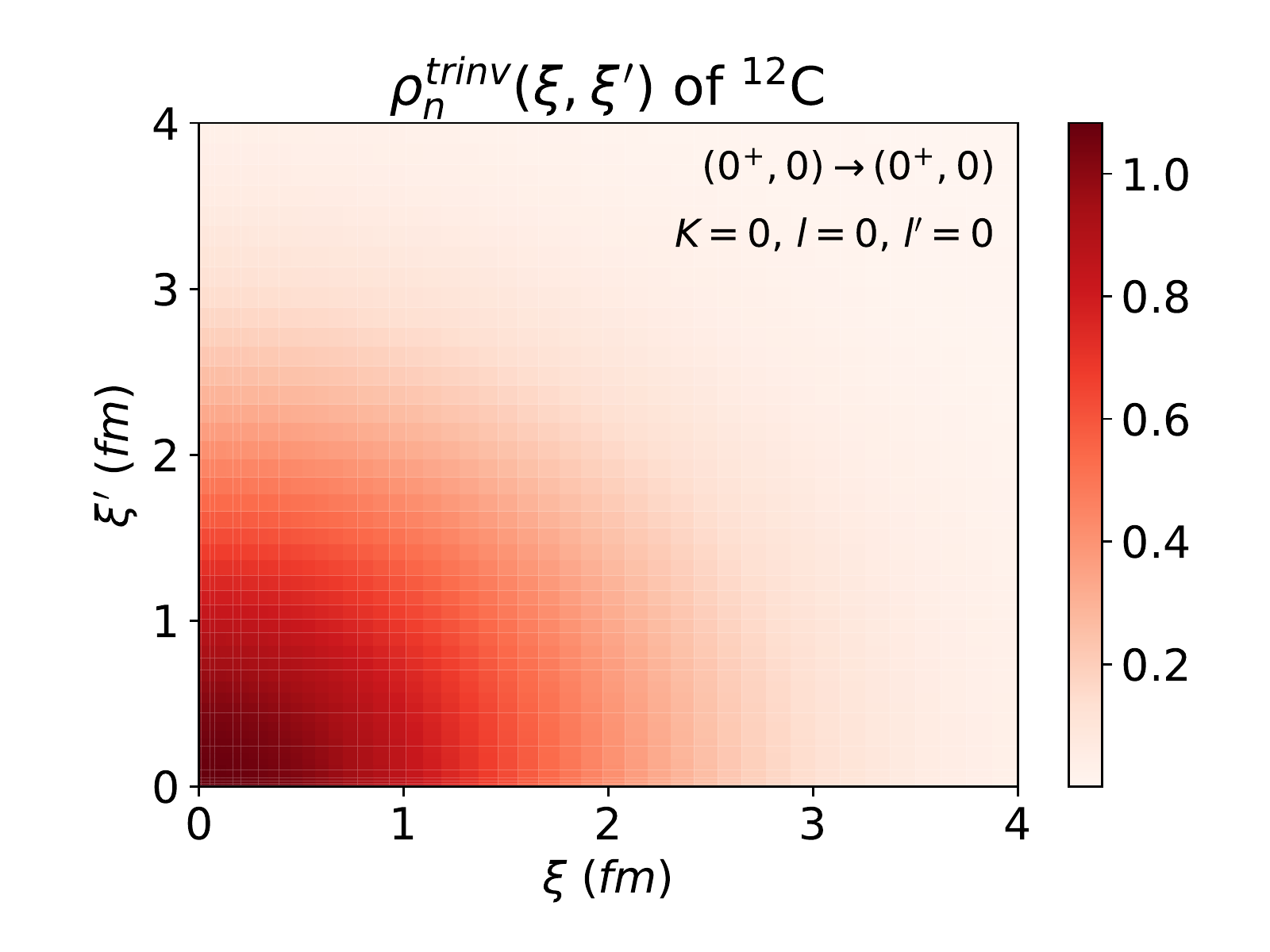}
 	\includegraphics[width=8.75cm]{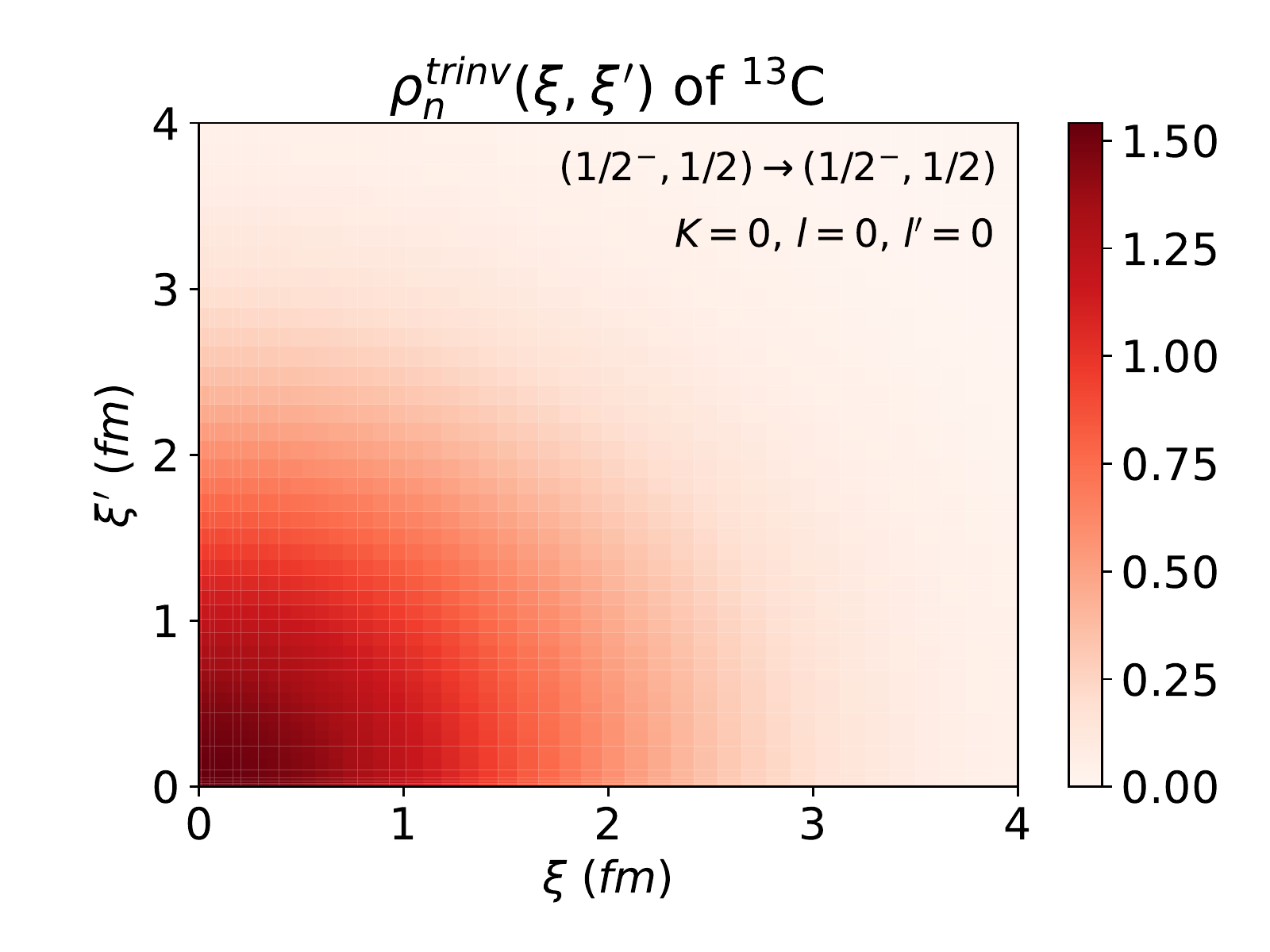}
 	\includegraphics[width=8.75cm]{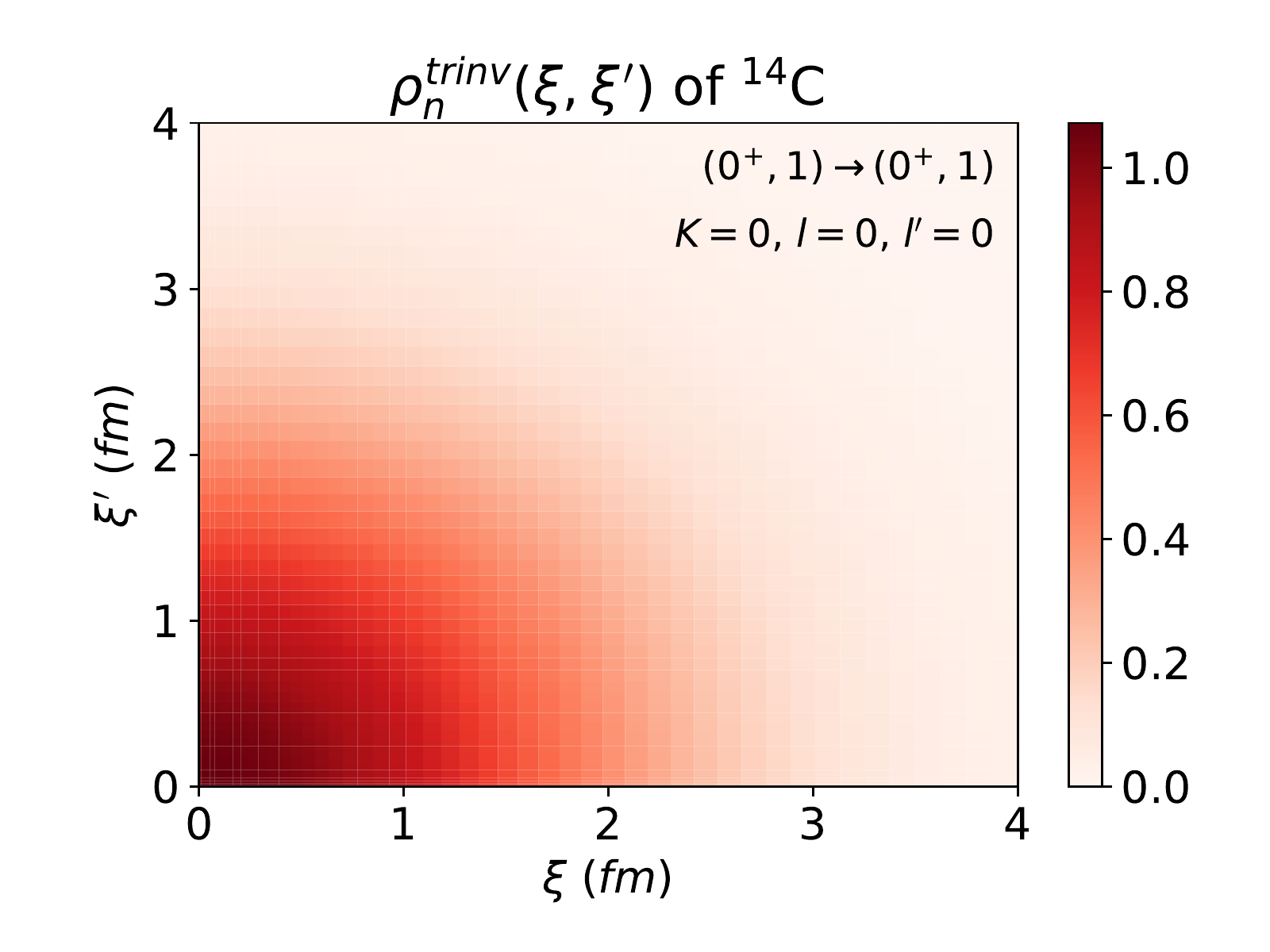}
 	\caption{\label{fig:densities_l0_CDB2K} Translation invariant nonlocal g.s. neutron densities for $^{10,11,12}\text{C}$ with an $N_\mathrm{max}=8$ model space and $^{13,14}\text{C}$ with an $N_\mathrm{max}=6$ model space, obtained within the NCSM. The $l=l'=0$ partial wave component is plotted. Densities were calculated using the CDB2K interaction with an oscillator frequency of $14 \ \text{MeV}$.}
\end{figure*}

\begin{figure*}
 	\includegraphics[width=8.75cm]{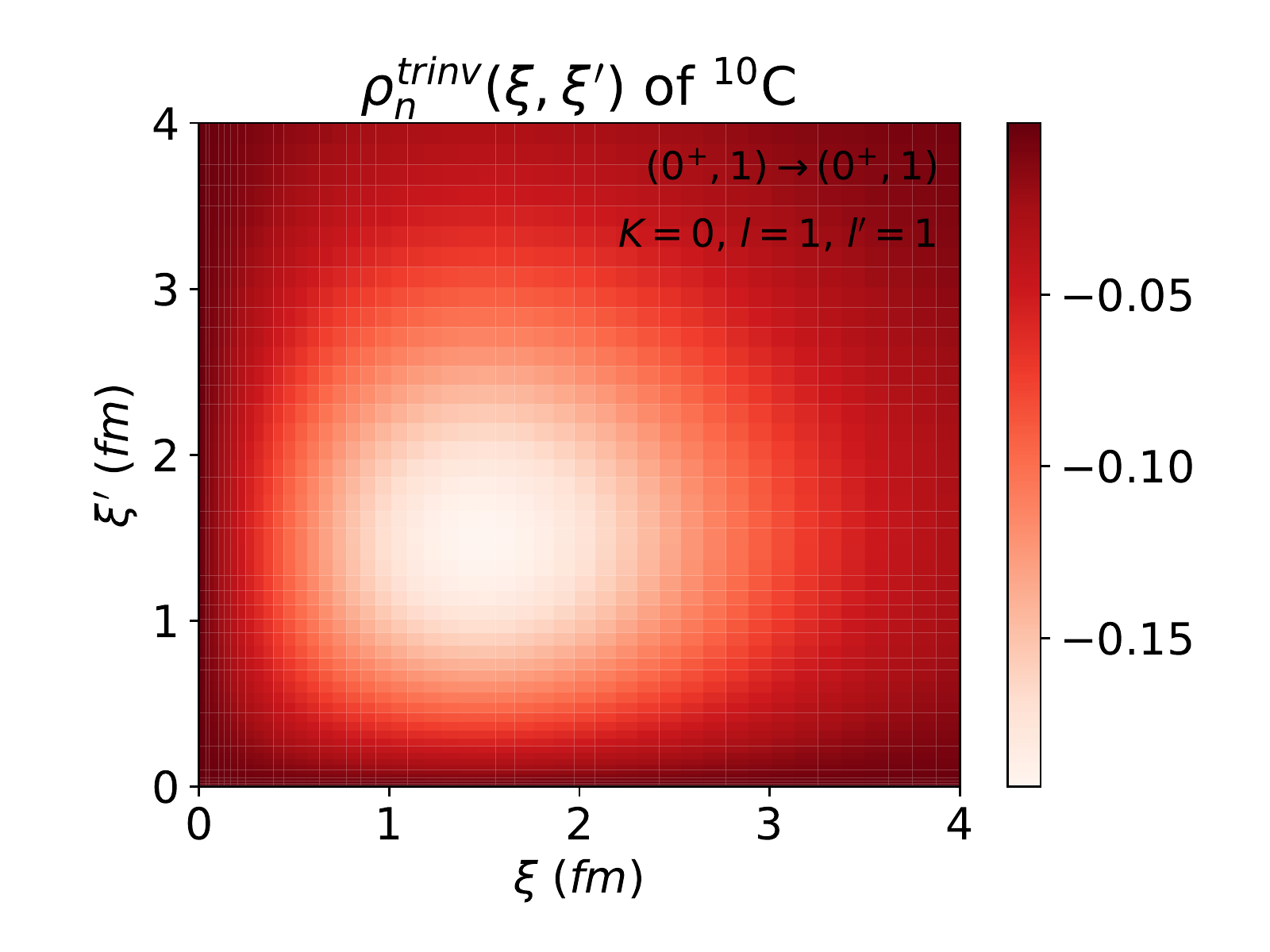}
 	\includegraphics[width=8.75cm]{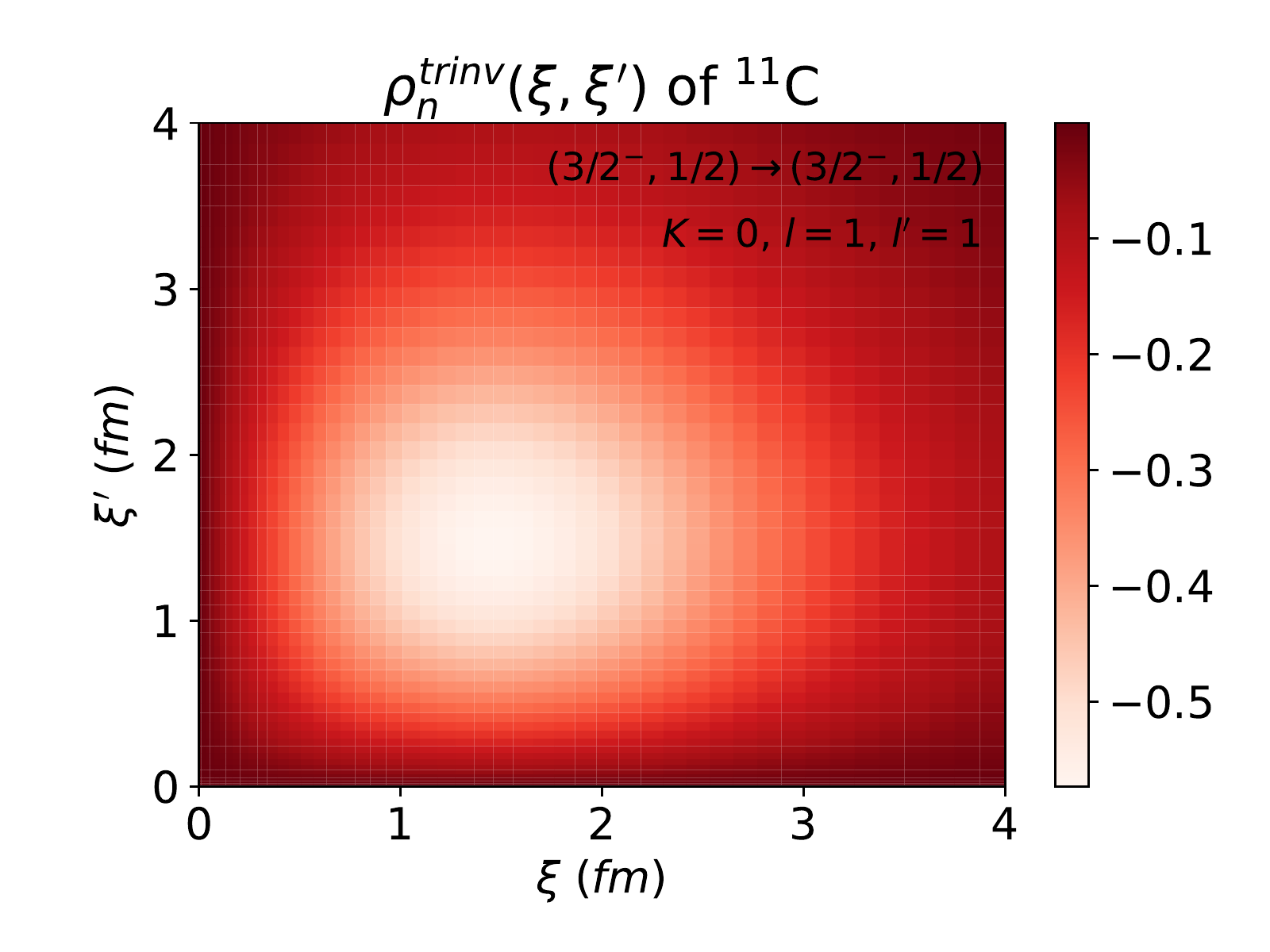}
 	\includegraphics[width=8.75cm]{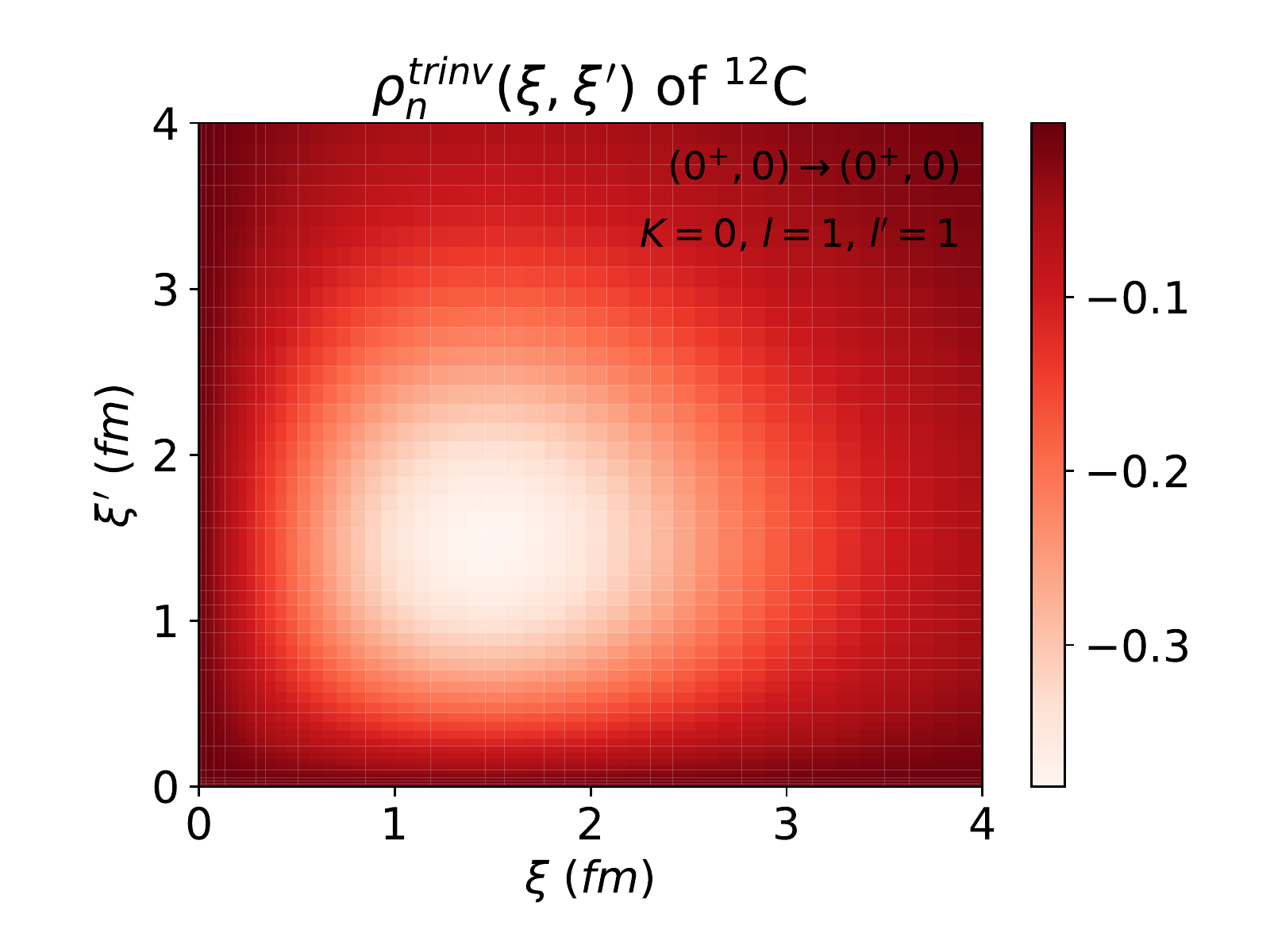}
 	\includegraphics[width=8.75cm]{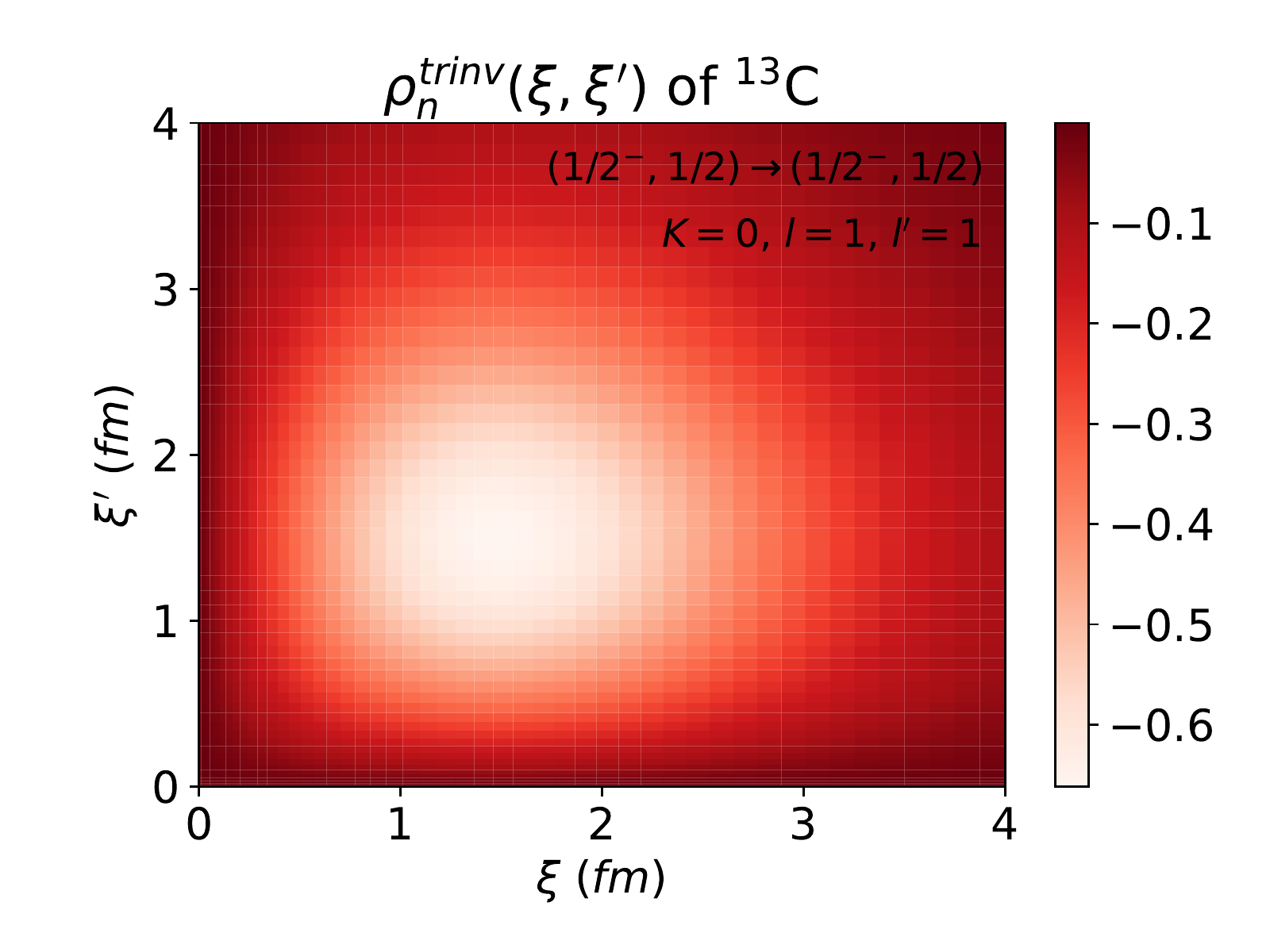}
 	\includegraphics[width=8.75cm]{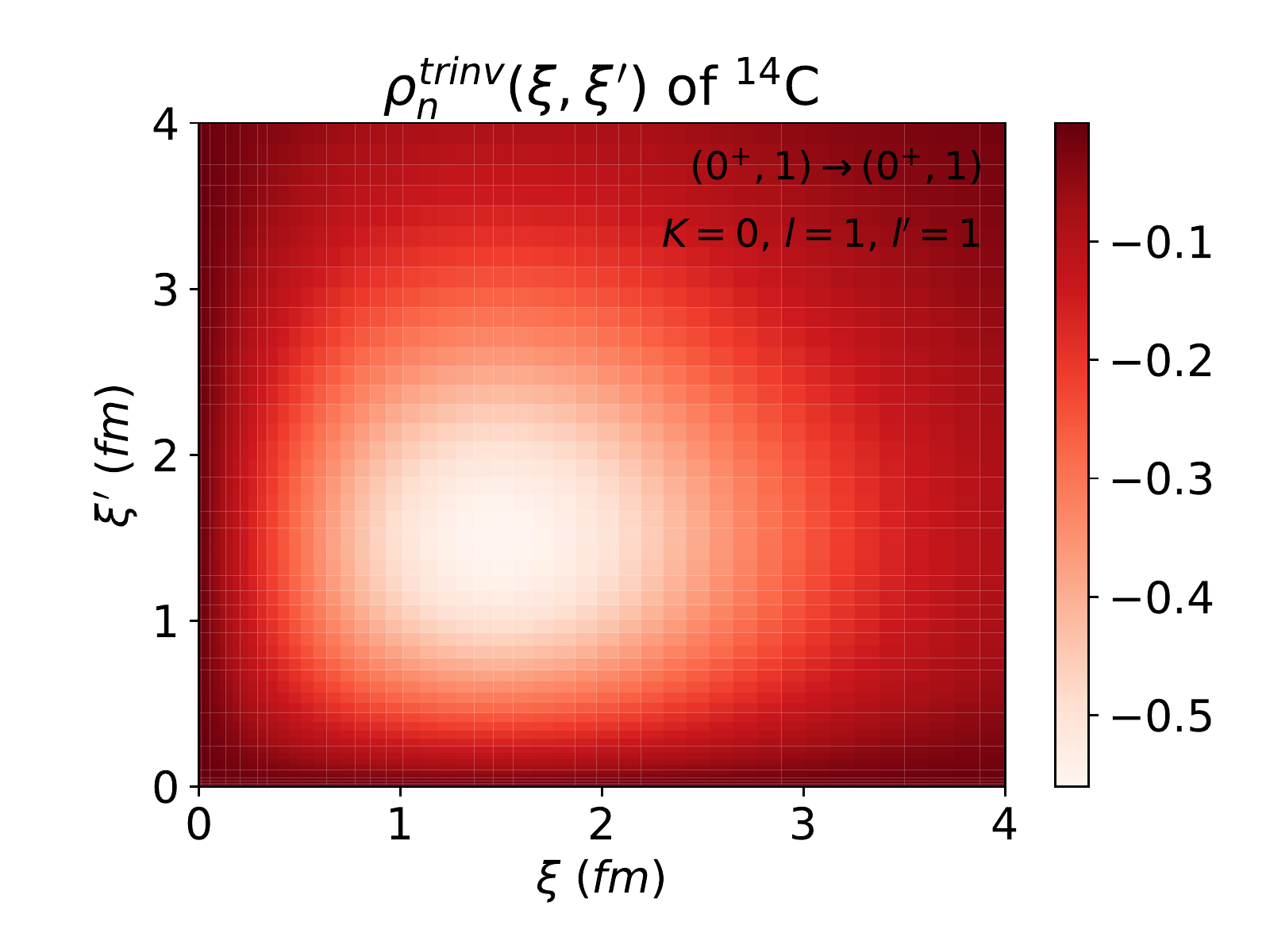}
 	\caption{\label{fig:densities_l1_CDB2K} Translation invariant nonlocal g.s. neutron densities for $^{10,11,12}\text{C}$ with an $N_\mathrm{max}=8$ model space and $^{13,14}\text{C}$ with an $N_\mathrm{max}=6$ model space, obtained within the NCSM. The $l=l'=1$ partial wave component is plotted. Densities were calculated using the CDB2K interaction with an oscillator frequency of $14 \ \text{MeV}$.}
\end{figure*}

\begin{figure}
 	\includegraphics[width=0.45\textwidth]{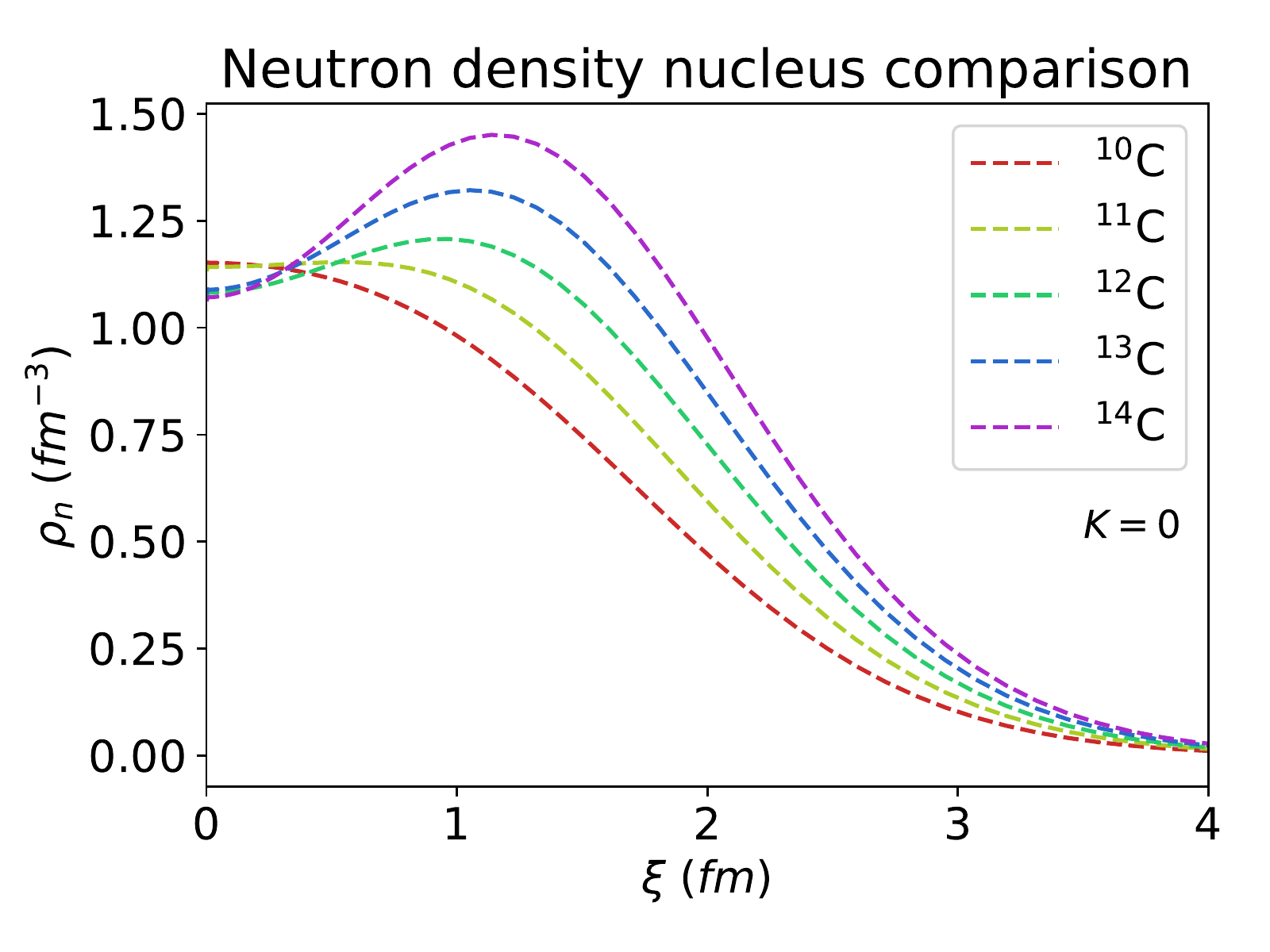}
 	\includegraphics[width=0.45\textwidth]{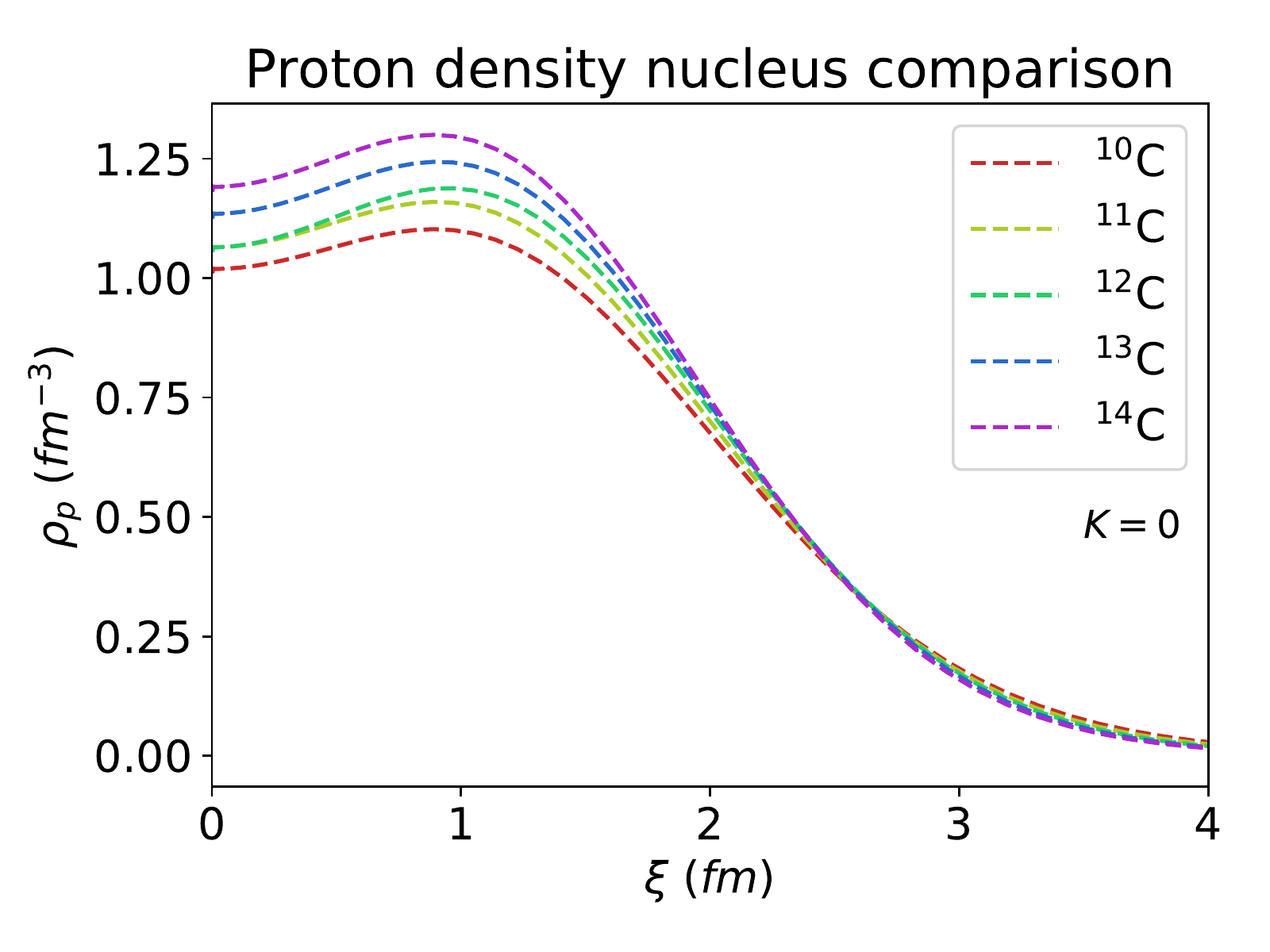}
 	\caption{\label{fig:local_nucleus_comp_CDB2K} Translation invariant local g.s. neutron (top) and proton (bottom) densities for $^{10,11,12}\text{C}$ computed with an $N_\mathrm{max}=8$ model space and $^{13,14}\text{C}$ computed with an $N_\mathrm{max}=6$ model space. The local density partial wave component shown is for zero transition momentum ($K=0$). Densities were calculated using the CDB2K interaction with an oscillator frequency of $14 \ \text{MeV}$.}
\end{figure}

\begin{figure*}
 	\includegraphics[width=8.75cm]{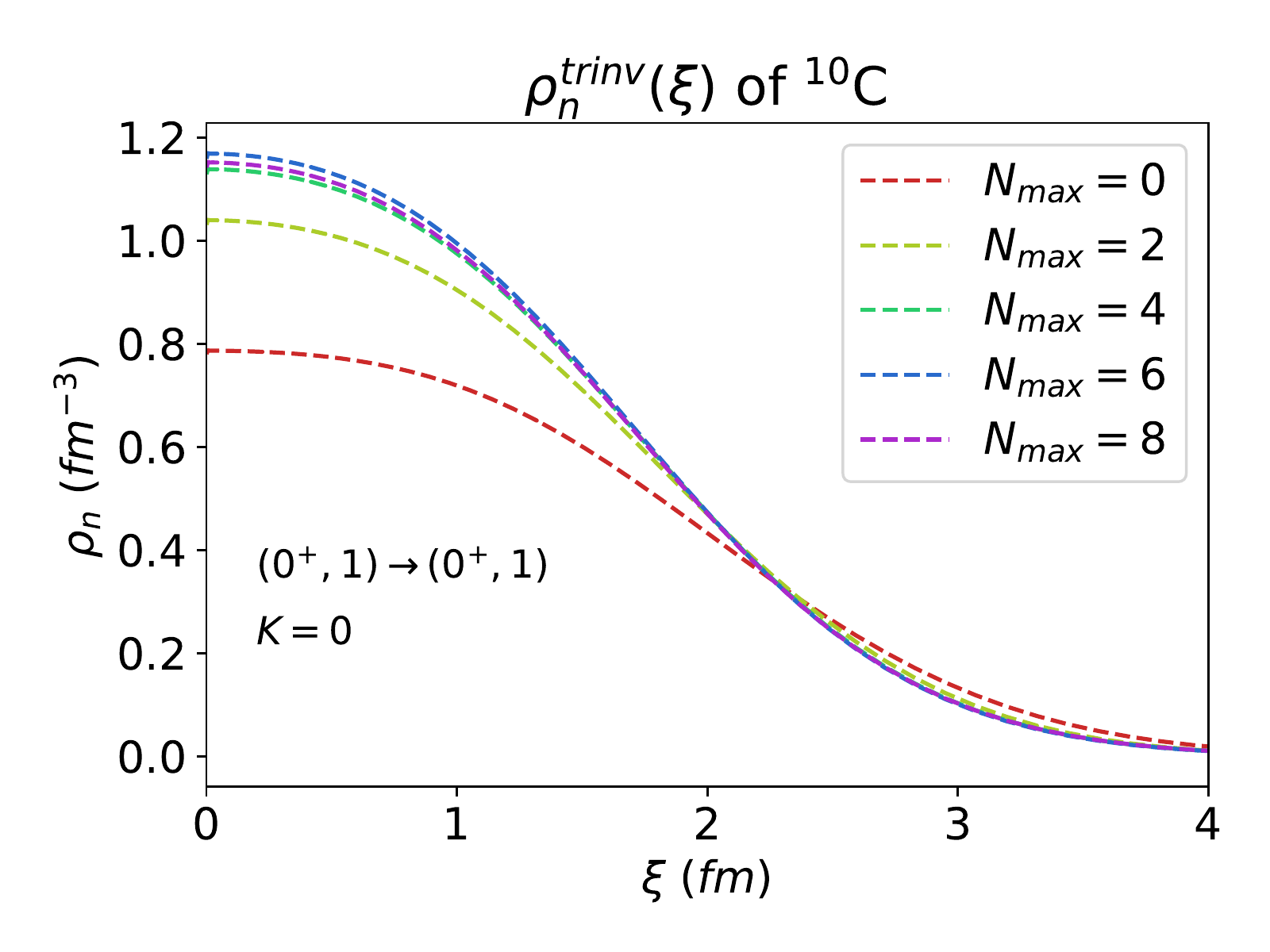}
 	\includegraphics[width=8.75cm]{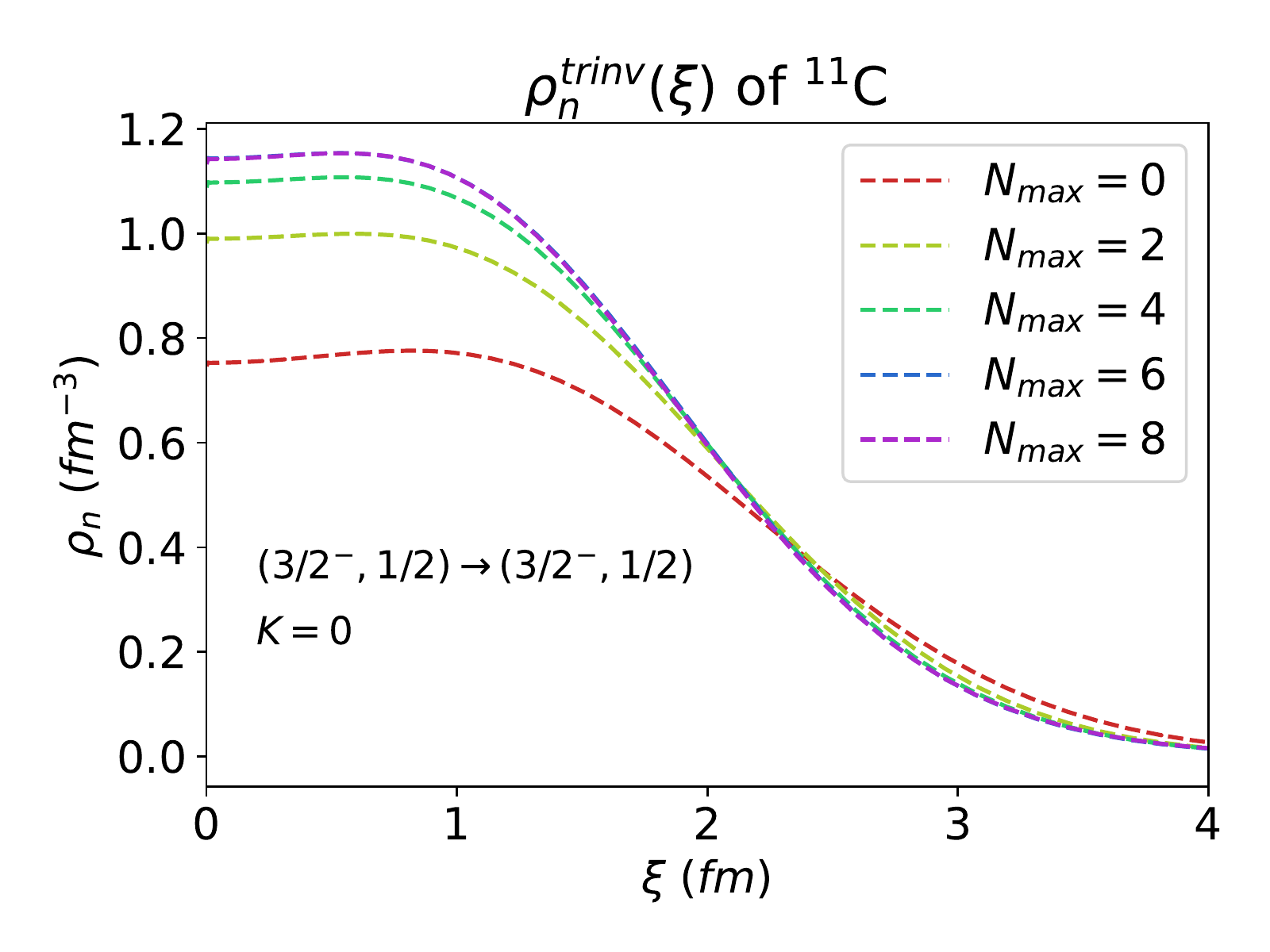}
 	\includegraphics[width=8.75cm]{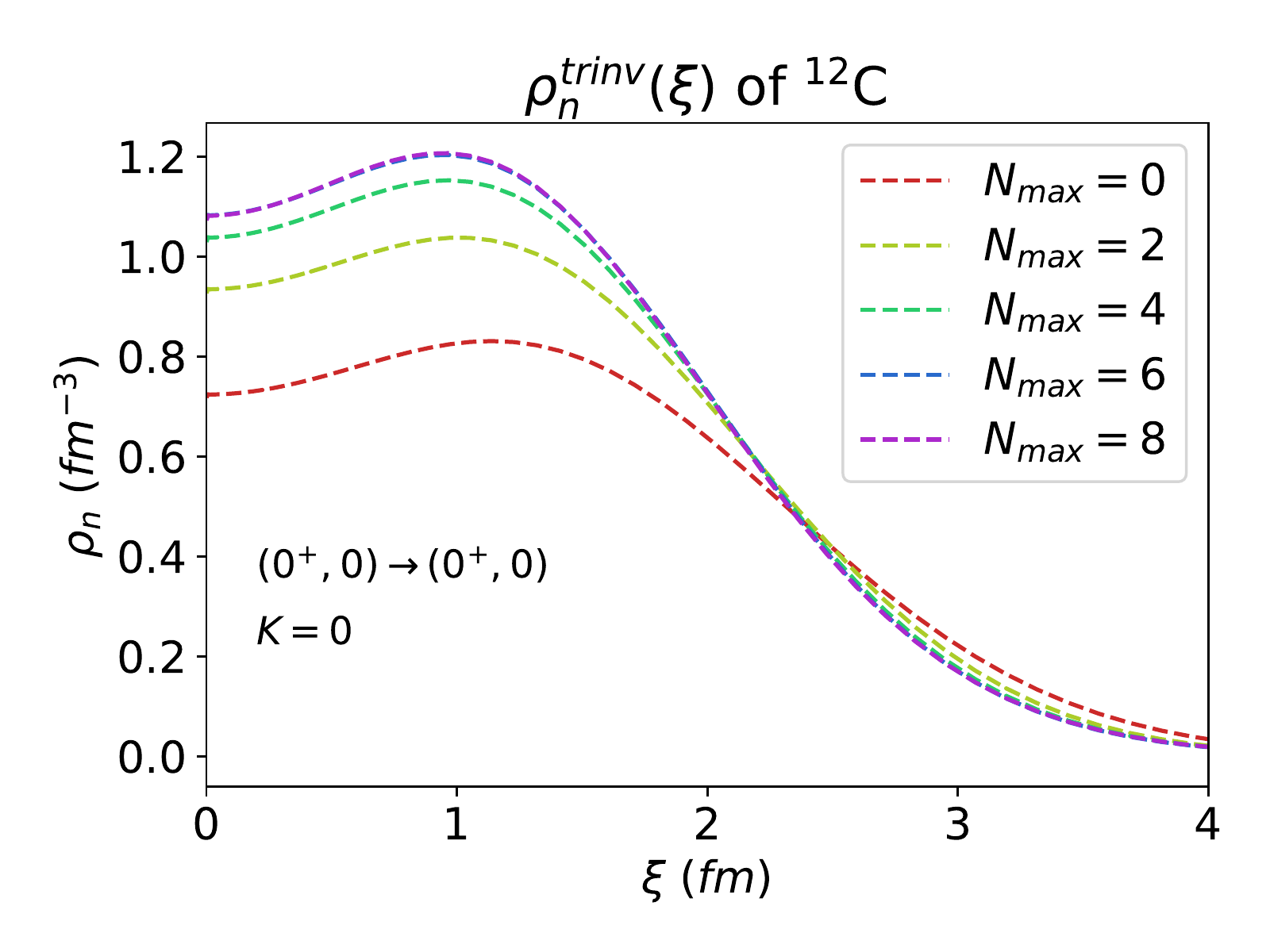}
 	\includegraphics[width=8.75cm]{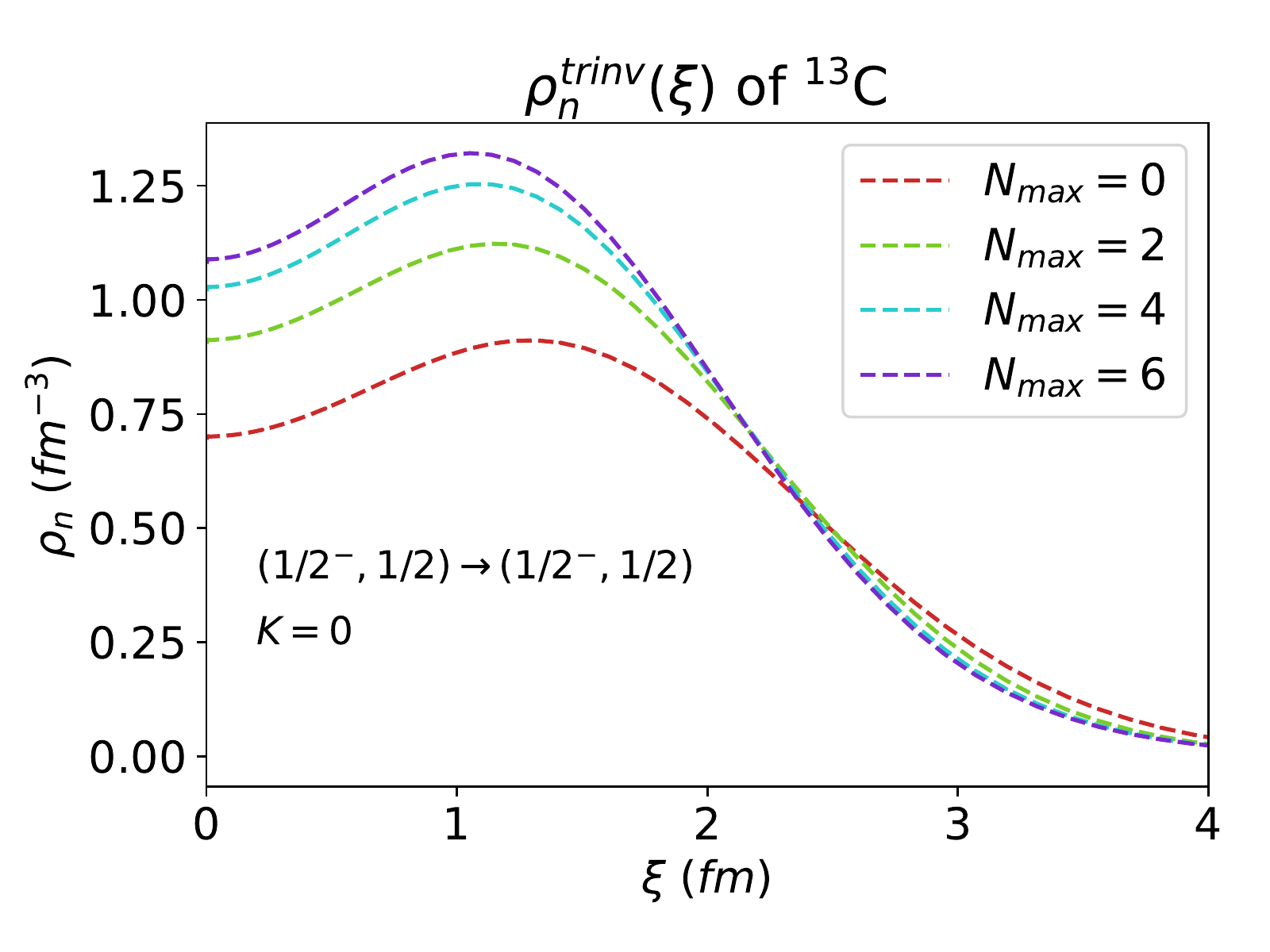}
 	\includegraphics[width=8.75cm]{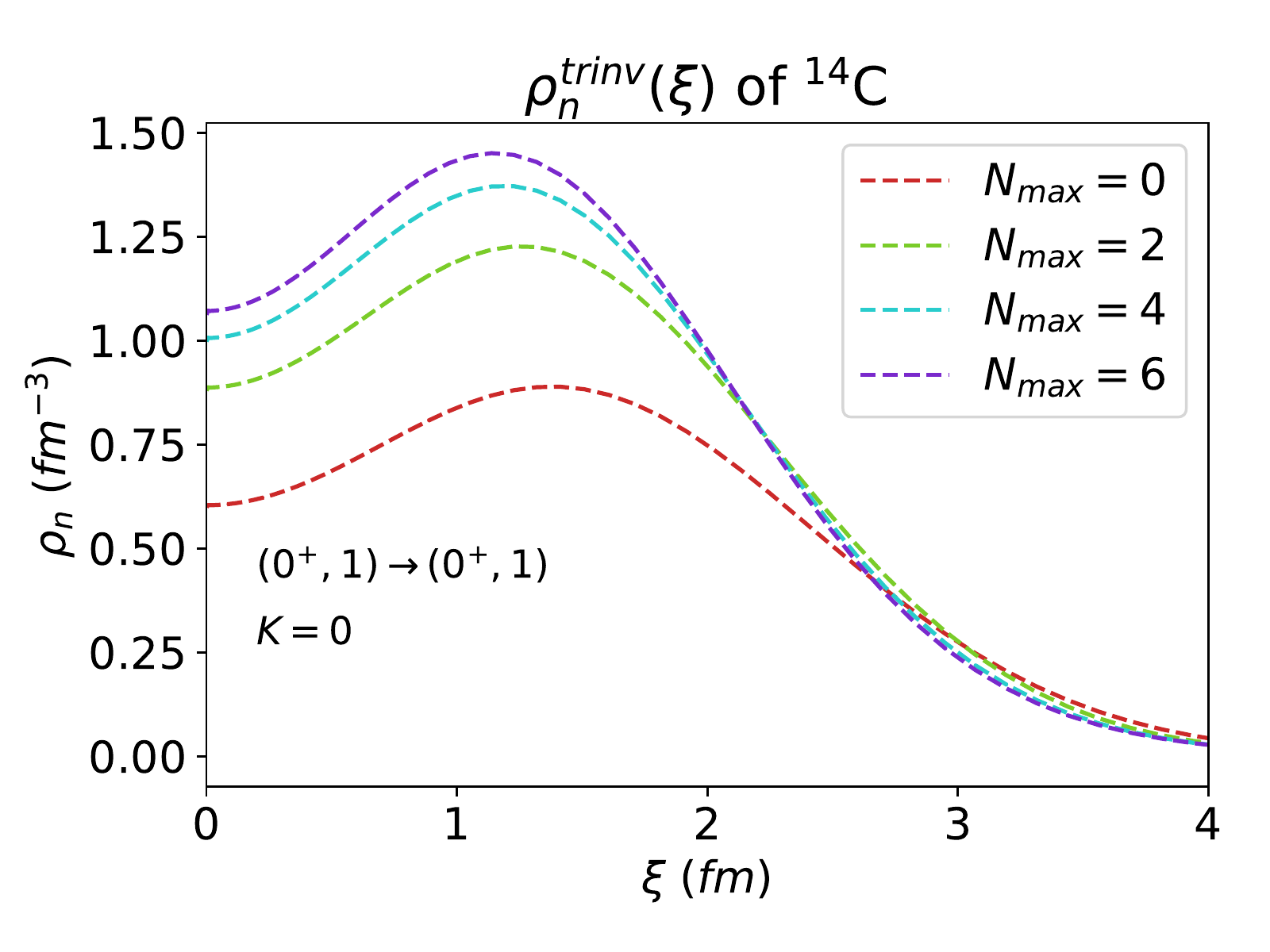}
 	\caption{\label{fig:local_densities_CDB2K} Translation invariant local g.s. neutron densities for $^{10,11,12}\text{C}$ computed with $N_\mathrm{max}=0-8$ models spaces and $^{13,14}\text{C}$ computed with $N_\mathrm{max}=0-6$ model spaces. The local density partial wave component shown is for zero transition momentum ($K=0$). Densities were calculated using the CDB2K interaction with an oscillator frequency of $14 \ \text{MeV}$.}
\end{figure*}
In addition to the aforementioned properties, with the $A$--nucleon eigenstates obtained in the NCSM it is possible to compute various density quantities. In this section we present calculations of the translation invariant nonlocal one--body nuclear densities for the range of carbon isotopes, utilizing the approach outlined in Refs.~\cite{Phys.Rev.C.97.034619,arXiv:2109.04017}. The translation invariant nonlocal density, written in partial wave components of $K$, $l$, and $l'$, is then given by
\begin{equation}\label{eq:trinv_nlocdens}
\begin{split}
& \rho^{fi}_{K l l'}(\vec{\xi},\vec{\xi}\,') = \frac{1}{\hat{J}_f} \sum (J_i M_i K k \vert J_f M_f) \bigg( Y^*_{l}(\hat{\xi}) Y^*_{l'}(\hat{\xi}\,') \bigg)^{(K)}_{k} \\
& \qquad \times R_{nl}(\xi) R_{n'l'}(\xi') \ (-1)^{l_1 + l_2 + K + j_2 + \frac{1}{2}} \ \hat{j}_1 \hat{j}_2 \hat{K} \\
& \qquad \times \begin{Bmatrix} j_1 & j_2 & K \\ l_2 & l_1 & \frac{1}{2} \end{Bmatrix} \ \big( M^K \big)^{-1}_{n_1 l_1 n_2 l_2 n l n' l'} \\
& \qquad \times \bigg[ \frac{-1}{\hat{K}} {}_{SD}\Big\langle A \lambda_f J_f \Big\vert \Big\vert \Big( a^{\dagger}_{n_1 l_1 j_1} \tilde{a}_{n_2 l_2 j_2} \Big)^{(K)} \Big\vert \Big\vert A \lambda_i J_i \Big\rangle{}_{SD} \bigg],
\end{split} 
\end{equation}
where the definition of the matrix $M^K$ is as in Ref.~\cite{Phys.Rev.C.97.034619}. The density is written in terms of the radial HO functions $R_{nl}(\xi)$, the spherical harmonics $Y_{l}(\hat{\xi})$ and the one--body density matrix elements obtained in a second quantization scheme. Due to the trivial antisymmetrization procedure, the NCSM eigenstates are obtained in the HO SD basis and are thus contaminated by ground state c.m. motion. The relative coordinates $\xi$ and $\xi'$, which measure the nucleon positions with respect to the c.m. coordinate $\vec{R}$, are then employed to analytically remove the spurious c.m. motion from the NCSM eigenstates, producing a translation invariant quantity.

For density calculations involving the isotopes $^{10-12}\text{C}$, the model space is set at $N_\mathrm{max}=8$ while for the  $^{13,14}\text{C}$ isotopes, the model space is set at $N_\mathrm{max}=6$. All densities presented in this section have been computed using an oscillator frequency of $14 \ \text{MeV}$ and the CDB2K interaction. We choose to primarily present the neutron densities as the proton distributions vary only slightly over the range of carbon isotopes. In all figures, the angular components of the density have been omitted and only the radial distributions are presented. In Fig.~\ref{fig:densities_l0_CDB2K}, we present the translation invariant nonlocal neutron densities for the carbon isotopes considered. The density itself is computed as a series of partial waves, for which we show the $l=l'=0$ component as this is generally the dominant partial wave contribution. One can see that the profile of the density is quite similar across all nuclei; the distribution being maximal near $\xi=\xi'=0$ and rapidly tapering off with increasing $\xi$ and $\xi'$. However, the density does vary substantially in magnitude for the even--odd nuclei, i.e. $^{11,13}\text{C}$, as compared to the even--even systems of $^{10,12,14}\text{C}$. In fact, for the even--even systems, the magnitude of the density remains quite similar despite the increasing number of neutrons. In Fig.~\ref{fig:densities_l1_CDB2K}, we present similar results, however this time for the $l=l'=1$ partial wave component of the density distribution. For this partial wave, the global maxima of magnitude of the density distribution is no longer located at $\xi=\xi'=0$ and is instead located around $\xi=\xi' \sim 1.5 \ \text{fm}$. Looking first at the $^{10,12,14}\text{C}$ isotopes, one notices that there is a significant enhancement in the magnitude of this component with increasing neutron number. Comparing to the isotopic behavior of the $l=l'=0$ partial wave, which remained almost unchanged with respect to the neutron number, the magnitude of the $l=l'=1$ component approximately triples going from $^{10}\text{C}$ to $^{14}\text{C}$. Furthermore, the magnitudes of the $^{11,13}\text{C}$ partial waves, which are notably larger than in the $^{10,12}\text{C}$ systems, are more comparable to the results from $^{14}\text{C}$.

In Fig.~\ref{fig:local_nucleus_comp_CDB2K}, we present a comparison of the local neutron (top plot) and proton (bottom plot) densities for all of the studied carbon isotopes. Note that the local density distribution is expressed in terms of the nonlocal form as 
\begin{equation}
\rho^{fi}_{K}(\vec{\xi}) = \sum_{l l'} \rho^{fi}_{K l l'}(\vec{\xi},\vec{\xi}\,') \Big\vert_{\xi=\xi'} \ .
\end{equation}
The normalization of the local proton and neutron densities is to the proton number $Z$ and neutron number $N$, respectively. For the local densities, we present the partial waves corresponding to zero transitions momentum ($K=0$), and in this case, we integrate the angular component analytically since the $Y^{0}_0(\hat{\xi})$ spherical harmonic trivially reduces to $\frac{1}{\sqrt{4\pi}}$. The angular factor of $\sqrt{4\pi}$ is included in all figures with the local density. Looking at the neutron densities for the various carbon isotopes, we see the following: (i) $^{10,11}\text{C}$ appear to be maximal at $\xi=0$ and only decay, though $^{11}\text{C}$ extends substantially further (ii) $^{12-14}\text{C}$ reach their maximal value at some non--zero $\xi$, in fact, the displacement increases with increasing neutron number. This is naturally explained in terms of the partial wave components of the nonlocal density shown in Fig.~\ref{fig:densities_l0_CDB2K} and Fig.~\ref{fig:densities_l1_CDB2K}, as those nuclei which have larger contributions from the $l=l'=1$ partial wave should see a more extended local density. As discussed prior, both $^{11,13}\text{C}$ have significant $l=l'=1$ partial waves, as do $^{12,14}\text{C}$, so the greater extent and magnitude of the local densities compared to $^{10}\text{C}$ is sensible. Referring now to the local proton densities, as expected, we find that they do not significantly differ in structure across the range of carbon isotopes.

Lastly, in Fig.~\ref{fig:local_densities_CDB2K}, we present $N_\mathrm{max}$ convergence plots obtained with the local densities. First looking at $^{10,11,12}\text{C}$, it is straightforward to see the emerging convergence trend, with increasing $N_\mathrm{max}$ model spaces providing more refined corrections to the density distribution. In these systems, the differences between the $N_\mathrm{max}=6$ and $N_\mathrm{max}=8$ distributions are almost negligible. In fact, the curves are exactly overlapping in the cases of $^{11,12}\text{C}$. Let us now consider the convergence plots of $^{13,14}\text{C}$, where the densities have only been obtained up to $N_\mathrm{max}=6$. For these systems, the difference between $N_\mathrm{max}=4$ and $N_\mathrm{max}=6$ is minutely larger than in the cases of the other nuclei. However, it is still reasonable to expect that, based on the convergence trends seen in the lighter carbon isotopes, the $N_\mathrm{max}=6$ density distribution would be in good agreement with the $N_\mathrm{max}=8$ distribution. Hence, it is safe to assume that these results are well converged in the NCSM.\\

\section{Conclusions}\label{sec:conclusions}

In this work, we have presented a systematic study of $^{10-14}\text{C}$ utilizing the \textit{ab initio} NCSM approach. We applied four NN interactions in the calculations, namely, the CDB2K, INOY, N\textsuperscript{3}LO and N\textsuperscript{2}LO\textsubscript{$opt$}, and comparison of NCSM results using these interactions have been carried out to determine best suited interaction. 
Low-lying energy spectra are investigated in the basis size of up to 10$\hbar\Omega$ for $^{10}$C and 8$\hbar\Omega$ for $^{11-14}$C. The triple--alpha structure of the $0_2^+$ state in $^{12}$C is obtained at high excitation energy even in the $N_\mathrm{max}=8$ basis space calculation. 
We note that entire energy spectrum is calculated at that optimal frequency which is obtained for g.s., thus, it is expected that convergence of this state would be obtained at different optimal frequency.
The energy of both the $T=0$ and $T=1$ $1^+$ states of $^{12}$C from \textit{ab initio} calculations are consistent with experiment. Regarding the energy of the ground state and some excited states, we generally find that the INOY interaction provides the best description of the energies. This confirms that inclusion of a short-range nonlocality in the NN interaction fitted to reproduce properties of the bound three-nucleon system ($^3$H and $^3$He) \cite{INOY,nonlocal,Doleschall} does explain some of the many–nucleon force effects.

We have calculated electromagnetic properties and compared these calculations with the available experimental data. \textit{Ab initio} NCSM results are consistent with experimental data except for $B(E2)$ value of $^{10}$C.  To understand the sensitivity of the point--proton radii with respect to the HO parameters, we have shown the dependence with respect to $\hbar\Omega$ and $N_{max}$ for $^{12}$C using INOY and N\textsuperscript{3}LO interactions. The N\textsuperscript{3}LO interaction adequately describes the radii of carbon isotopes when compared to experiment. It is also worth noting that the optimal frequency for determining the point--proton radii is smaller than the optimal frequency of the many--body calculation. Lastly, for the first time, we have reported the translation invariant one--body nuclear densities for $^{10-14}$C and their behavior with respect to $N_\mathrm{max}$.

\vspace*{0.3cm}
\section*{ACKNOWLEDGMENTS}
We acknowledge the research grant CRG/2019/000556 from SERB (India).  P.C. acknowledges financial support from the MHRD (Government of India) for her Ph.D. thesis work. P.N. acknowledges support from the NSERC Grant No. SAPIN-2022-00019. TRIUMF receives federal funding via a contribution agreement with the National Research Council of Canada. We would like to thank Christian Forss{\'e}n for making the pAntoine code available.

\bibliographystyle{utphys}
  \bibliography{references}


\end{document}